\documentclass[aps,prd,twocolumn,nofootinbib,superscriptaddress,balancelastpage]{revtex4}%letterpaper

\usepackage[T1]{fontenc} % if needed
\usepackage[utf8]{inputenc}
\usepackage{graphicx}
\usepackage{graphics}
\usepackage[mathscr]{eucal}
\usepackage{amssymb} 
\usepackage{amsmath}
\usepackage{xspace}
\usepackage{listings}
\usepackage[normalem]{ulem}
\usepackage{autobreak}
\usepackage{xcolor}
\usepackage{bm}         % for bold math characters in a math einvironment
\usepackage{array}
\usepackage{multirow}   % analogue to \multicolumn
\usepackage{booktabs}   % for fancy looking tables
\usepackage{mleftright}
\usepackage{lipsum}

\usepackage{bbold}
\usepackage{amsmath,amscd}
\usepackage{slashed}
\usepackage{placeins}
\usepackage{soul}
\usepackage{braket}
\usepackage{empheq}
\usepackage[most]{tcolorbox}
\usepackage{mathtools,slashed}
\usepackage{dsfont}
\usepackage[makeroom]{cancel}
\usepackage{multirow}
\usepackage{amssymb}

\usepackage{dcolumn}% Align table columns on decimal point
\usepackage{physics}
\usepackage{makecell} 
\renewcommand\arraystretch{1}
\usepackage{hyperref}
\hypersetup{colorlinks=true, linkcolor=blue, urlcolor=blue, citecolor=blue}

\usepackage[capitalise]{cleveref}
\crefname{section}{Sec.}{Secs.}
\crefname{table}{Tab.}{Tabs.}
\crefname{figure}{Fig.}{Figs.}
\crefname{equation}{Eq.}{Eqs.}
\crefname{appendix}{Appendix\ }{Appendix\ }
\usepackage{comment}

\renewcommand{\[}{\left[}

\definecolor{ForestGreen}{rgb}{0.13, 0.55, 0.13}
\definecolor{Mulberry}{rgb}{0.77, 0.29, 0.55}
\definecolor{bostonuniversityred}{rgb}{0.8, 0.0, 0.0}
\definecolor{amber}{rgb}{1.0, 0.49, 0.0}

\newcommand\varpm{\mathbin{\vcenter{\hbox{%
  \oalign{\hfil$\scriptstyle\hspace{-0.1ex}+\hspace{-0.1ex}$\hfil\cr
          \noalign{\kern-.5ex}
          $\scriptscriptstyle({-})$\cr}%
}}}}

\allowdisplaybreaks
\begin{document}

\title{Two-peak gravitational wave spectra from supercooled two-field phase transitions}

\author{Daniel Ara\'{u}jo}
\email{dfcoaraujo@lip.pt}
\affiliation{Laboratório de Instrumentação e Física Experimental de Partículas (LIP), Universidade do Minho, 4710-057 Braga, Portugal}
\affiliation{Departamento de F\'{i}sica, Escola de Ci\^{e}ncias, Universidade do Minho, 4710-057 Braga, Portugal}
\author{João Gonçalves}
\email{jppinog@lip.pt}
\affiliation{Laboratório de Instrumentação e Física Experimental de Partículas (LIP), Universidade do Minho, 4710-057 Braga, Portugal}
\author{Danny~Marfatia}
\email{dmarf8@hawaii.edu}
\affiliation{Department of Physics and Astronomy, University of Hawaii at Manoa, Honolulu, HI 96822, USA}

\begin{abstract}
We study the gravitational wave (GW) spectrum generated by supercooled first-order phase transitions in two-field conformal models. 
We construct three representative toy models that capture the essential features of these models, derive analytical criteria for the existence of two-step transitions,
and show how the two-field dynamics generate a genuine two-peak GW spectrum.
We identify regions of parameter space in which two-peak GW spectra are observable at LISA and the Einstein Telescope. 
By way of illustration, we apply our analysis to a realistic conformal dark SU(3) model.
\end{abstract}

\maketitle
%%%%%%%%%%%%%%%%%%%%%%%%%%%%%%%%%%%%%%%%%%%%%%%%%%%%%%%%%%
\section{Introduction}\label{sec:intro}
%%%%%%%%%%%%%%%%%%%%%%%%%%%%%%%%%%%%%%%%%%%%%%%%%%%%%%%%%%
Models with classical scale-invariance—a special case of conformal invariance—are well-known examples for the generation of strong gravitational wave (GW) signals from first-order phase transitions (FOPTs) because they generally predict a significant amount of supercooling~\cite{Jinno:2016knw,Jaeckel:2016jlh,Ellis:2020nnr,Kierkla:2022odc}. 
To date, most studies have focused on the one-field limit. Multi-field scenarios, which can generate a multi-peak GW spectrum, are underexplored.

Examples of two-peak GW spectra arise in nonconformal multi-Higgs extensions. However, these scenarios predict that at least one of the peaks has an amplitude that falls below the sensitivity of planned observatories~\cite{Morais:2019fnm,Liu:2023sey}. 
Observable two-peak spectra were obtained in the context of flavor models~\cite{Greljo:2019xan,Fabri:2025fsc}, although the two-peak spectra are determined by modeling the different phase FOPTs with independent single-field potentials.
 In contrast, we consider the full two-field dynamics, which due to the large degree of supercooling expected in conformal models, can produce distinct peaks within reach of future GW observatories. 
  Thus, these models offer a more interesting signature compared to the two-peak spectra of nonconformal and flavor models.

To keep our analysis as generic as possible, we construct a set of toy models that capture the essential features of conformal models. Instead of the purely numerical treatment found in the literature, our approach has the added benefit of being analytically tractable. We derive analytical criteria for the existence of two-step FOPTs, and develop
semianalytical methods to evaluate the tunneling action by extending the results of Ref.~\cite{Pascoli:2026tuu}, originally derived for one-field models, to the two-field case. 

The paper is organized as follows. In Sec.~\ref{sec:toy_models} we present three toy models to study the two-field dynamics and obtain conditions for a two-step transition to occur. In Sec.~\ref{sec:phase_analytical}, we describe our semi-analytical approach to write the thermal potential in polynomial form and compute the bounce action. In Sec.~\ref{sec:GWs} we outline our numerical procedure for computing the GW spectra, and in Sec.~\ref{sec:num_results}, we present and discuss our numerical results. In Sec.~\ref{sec:summ} we summarize our main conclusions. 

%%%%%%%%%%%%%%%%%%%%%%%%%%%%%%%%%%%%%%%%%%%%%%%%%%%%%%%%%%
\section{Toy models}\label{sec:toy_models}
%%%%%%%%%%%%%%%%%%%%%%%%%%%%%%%%%%%%%%%%%%%%%%%%%%%%%%%%%%
 We construct conformal toy models with two real scalars, $\varphi_1$ and $\varphi_2$, and the tree-level potential,
\begin{equation}\label{eq:tree_level}
    V_0 = \frac{\lambda_1}{4} \varphi_1^4 + \frac{\lambda_{12}}{4} \varphi_1^2 \varphi_2^2 + \frac{\lambda_2}{4} \varphi_2^4\,.
\end{equation}
The Coleman-Weinberg (CW) mechanism breaks the classical conformal symmetry radiatively and generates two nontrivial vacuum expectation values (vevs) $v_1$ and $v_2$. We consider three realizations of the corresponding one-loop CW potential, that define three distinct toy models. Model TM1 consists of two vector fields, $X_1$ and $X_2$; model TM2 contains three vector fields, $X_1$, $X_2$ and $X_3$; and model TM3 contains two vector fields, $X_1$ and $X_2$, plus one fermion field, $\psi_1$. The CW potentials for these toy models are given by 
\begin{equation}\label{eq:one1loop}
\begin{split}
    & V_{\rm CW} = \sum_{i} \frac{n_iM_i^4}{64\pi^2}\Big[ \ln\frac{M_i^2}{\mu^2} - a_i\Big]\,, \quad i = X_1,X_2,X_3,\psi_1\\
    &\mathrm{TM1}: M_{X_1} = g_1\varphi_1,\, M_{X_2} = g_2 \varphi_2,\,  \\
    &\mathrm{TM2}: M_{X_1} = g_1\varphi_1,\, M_{X_2} = g_2 \varphi_2,\, M_{X_3}=g_3\sqrt{\varphi_1^2+\varphi_2^2}\,,\\
    & \mathrm{TM3}: M_{X_1}=g_1\varphi_1,\,  M_{X_2}=g_2\varphi_2,\, M_{\psi_1}=y(\varphi_1 + \varphi_2) \,,
\end{split}
\end{equation}
where $\mu = 91~\mathrm{GeV}$ is the renormalization scale, $a=5/6$ for bosons and $a=3/2$ for fermions, $n_i=(-1)^{2s_i}Q_iN_i(2s_i+1)$ are the degrees of freedom, where $s$ is the particle's spin, and $Q= 1,2$ for uncharged particles and charged particles, respectively, and $N=1,3$ for uncolored particles and colored particles, respectively.
We do not consider models in which the CW term has contributions from scalar fields because we implicitly assume the scaling $\lambda_i \sim  g_i^4$, typically found in conformal models~\cite{Weinberg:1973am}, so that scalar contributions are subdominant to vector contributions for $g_i < 1$. 
For each toy model, we employ the tadpole equations in Appendix~\ref{app:tadpoles} as conditions on $\lambda_1$ and $\lambda_2$.

\begin{figure*}[t]
    \centering
    \includegraphics[width=0.65\textwidth]{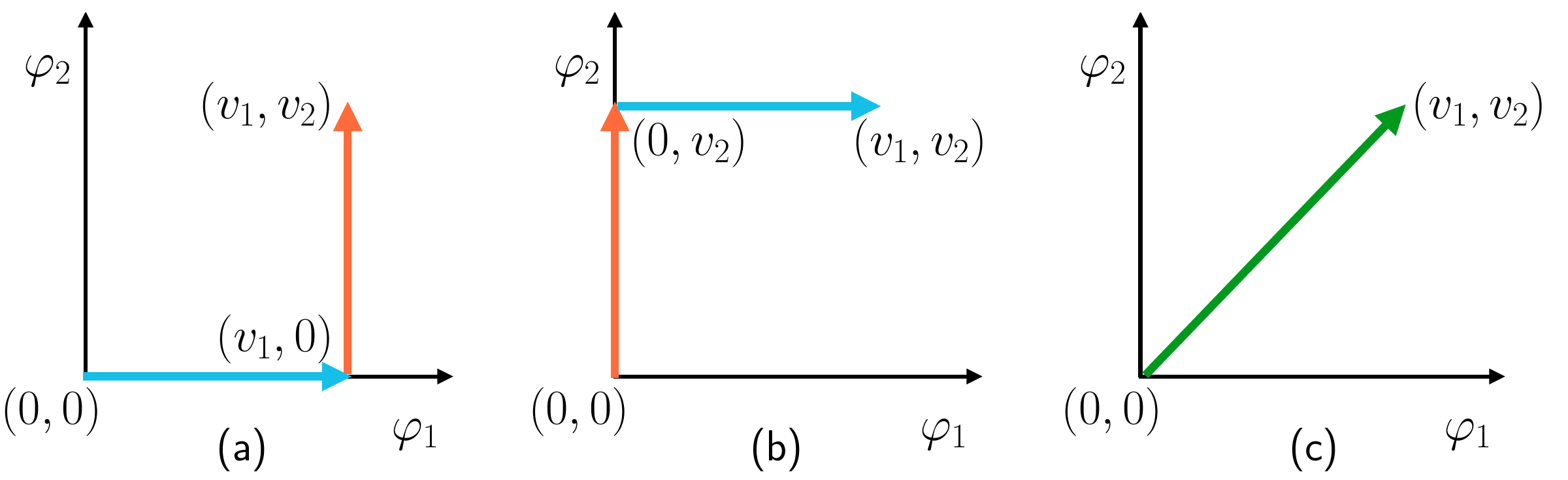}   
    \caption{Schematic representation of the phase transition patterns in our two-field toy models. Panels~(a) and~(b) illustrate the two-step transitions, $(0,0) \rightarrow (v_1,0) \rightarrow (v_1,v_2)$ and $(0,0) \rightarrow (0,v_2) \rightarrow (v_1,v_2)$, respectively. Panel~(c) illustrates the one-step transition, $(0,0) \rightarrow (v_1,v_2)$. 
}
    \label{fig:phase_twostep}
\end{figure*}

The general features of these toy models arise in ultraviolet complete Abelian or non-Abelian gauge models. Specifically, TM1 and TM3 are inspired by dark 
$\mathrm{U(1)^\prime}$ models~\cite{Ellis:2020nnr,Goncalves:2024lrk,Jinno:2016knw} while TM2 is inspired by a non-Abelian dark SU(3) model~\cite{Karam:2016rsz}. Note that these realistic models have a single gauge coupling. Thus, dark $\mathrm{U(1)^\prime}$ models correspond to TM1/TM3 with $g_1 = g_2$, and the dark SU(3) model corresponds to TM2 with $g_1=g_2=g_3$. Nevertheless, we first consider the general scenario in which each $M_i$ has an independent coupling, so as to isolate its contribution. Later, we discuss the realistic limit, as well as a fully realistic model.

An important criterion for the occurrence of a two-step FOPT is the existence at zero temperature, of extrema in different regions of field space. To apply the one-field approximation in the direction of each phase transition,
the potential must admit extrema at $(0,0)$, $(v_1,0)$ or $(0,v_2)$, and $(v_1,v_2)$. We focus on TM1 to classify the different extrema. Analogous calculations apply for TM2 and TM3. Using the tadpole equations, the one-loop field-dependent Hessian matrix is
\begin{equation}\label{eq:hessian_matrix}
    M_{\varphi}^2 = \begin{pmatrix}
        \dfrac{3g_1^4\varphi_1^2}{8\pi^2} - \lambda_{12}\varphi_2^2 & \lambda_{12}\varphi_1\varphi_2 \\
        \lambda_{12}\varphi_1\varphi_2 & \dfrac{3g_2^4\varphi_2^2}{8\pi^2}-\lambda_{12}\varphi_1^2
    \end{pmatrix}\,.
\end{equation}
Besides the trivial extremum at the origin, there are three 
nontrivial extrema that we can classify:
\begin{itemize}
    \item[1)] $(\varphi_1,\varphi_2)=(v_1,v_2)$. In this case, the mass eigenvalues are
    \begin{equation}
        m_{h_1,h_2}^2 = \frac{1}{2}(a+c\mp\sqrt{a^2+4b^2-2ac+c^2})\,, 
    \end{equation}
    where $a = (3g_1^4v_1)/8\pi^2 - \lambda_{12}v_2^2$, $b=\lambda_{12}v_1v_2$ and $c=(3g_2^4v_2)/8\pi^2 - \lambda_{12}v_1^2$. Although this expression is cumbersome, it provides useful insight in limiting cases. For $g_1,g_2 \gg |\lambda_{12}|$, the mixing can be neglected and the physical squared masses $m_{h_1,h_2}^2 \simeq 3g_2^4 v_2^2/8\pi^2$ are positive.  In the opposite limit, $|\lambda_{12}| \gg g_1,g_2$, we find $m_{h_1}^2 \simeq -\lambda_{12}(v_1^2 + v_2^2)$ and $m_{h_2}^2 = \mathcal{O}(g_1^4 + g_2^4)$. In this regime, $\lambda_{12} < 0$ is required to ensure that $m_{h_1}^2>0$. More generally, for both squared masses to be positive requires that
    \begin{equation}\label{eq:general_cond}
        \lambda_{12} < \frac{3}{8\pi^2}\left(\frac{g_1^4 g_2^4 v_1^2 v_2^2}{g_1^4 v_1^4 + g_2^4 v_2^4}\right)\,.
    \end{equation} 
    For typical values of the gauge couplings, $g_1,g_2 = 0.1-1$, and vevs, $v_1,v_2 = 150 - 10^4~\mathrm{GeV}$, this condition allows $\lambda_{12}$ to be positive, but it cannot be much larger than $\mathcal{O}(0.01)$.
    
    \item[2)] $(\varphi_1,\varphi_2)=(0,v_2)$.  At this extremum, the mass eigenvalues are
    \begin{equation}\label{eq:masses_mix}
    m^2_{h_1} = -\lambda_{12} v_2^2\,,\quad m^2_{h_2} = \frac{3g_2^2v_2^2}{8\pi^2}\,.
    \end{equation}
    Thus, for $(0,v_2)$ to be a local minimum, $\lambda_{12} < 0$. Although a negative $\lambda_{12}$ is favored for an intermediate minimum to lie on the axis, this is not strictly required. At finite temperature, thermal corrections to the scalar masses scale as $g_2^2T^2$ and can be sizable enough to render $m_{h_1}^2$ positive even if $\lambda_{12} > 0$. 

    \item[3)] $(\varphi_1,\varphi_2)=(v_1,0)$.
    This is equivalent to 2) under the transformation, $v_2 \rightarrow v_1$, $g_2 \rightarrow g_1$ and $\lambda_1 \rightarrow \lambda_2$. 
\end{itemize}

In summary, positivity of the squared masses at $(v_1,v_2)$ constrains $\lambda_{12}$ to values not much larger than $\mathcal{O}(0.01)$ for typical model parameters, while the existence of a minimum on either axis, $(0,v_2)$ or $(v_1,0)$, requires \mbox{$\lambda_{12} < 0$} at zero temperature. 

Since a two-step FOPT requires the existence of an intermediate phase, it is instructive to investigate how the minima along the axes depend on the model parameters. By projecting the potential in the $\varphi_1 = 0$ direction, and using the tadpole conditions, we obtain
\begin{equation}\label{eq:pot_dir1}
    \begin{aligned}
    V_{\varphi_2} = &-\frac{\lambda_{12} \varphi_2^4 v_1^2}{8 v_2^2}  \\
    &-\frac{3 g_2^4 \varphi_2^4}{128 \pi ^2} \Bigg(2 \ln \left(\frac{g_2^2 v_2^2}{\mu ^2}\right) - 2 \ln \left(\frac{g_2^2 \varphi_2^2}{\mu ^2}\right)+1\Bigg)\,.
    \end{aligned}
\end{equation}
Here, $v_2$ is the vev corresponding to the minimum at $(v_1,v_2)$, and does not necessarily coincide with the location of the minimum along the $\varphi_2$ axis when $\varphi_1 = 0$. Solving the tadpole equation in Eq.~\eqref{eq:pot_dir1} for the location of the minimum gives
\begin{equation}\label{eq:vev_dir1}
    \expval{\varphi_2} = v_2 \exp\Bigg({\frac{4 \pi ^2 \lambda_{12} v_1^2}{3 g_2^4 v_2^2}}\Bigg)\,.
\end{equation}
These expressions illustrate the importance of $\lambda_{12}$ in determining the existence and location of a minimum on the $\varphi_2$ axis. First, they confirm a preference for $\lambda_{12} < 0$ because the first term in Eq.~\eqref{eq:pot_dir1} is positive, which can make the potential bounded from below. However, a positive $\lambda_{12}$ is possible if it is compensated by a sufficiently large gauge coupling; see Eq.~\eqref{eq:general_cond}. Second, Eq.~\eqref{eq:vev_dir1} shows that the location of the minimum on the $\varphi_2$  axis depends exponentially on $\lambda_{12}$. Consequently, sufficiently large negative values of $\lambda_{12}$ makes the minimum on the axis rapidly approach $\expval{\varphi_2}=0$. As such, without a minimum on the axis, a two-step transition is not possible and only the $(0,0)\rightarrow (v_1,v_2)$ transition is allowed. In contrast, as $\lambda_{12} \rightarrow 0$, the axis minimum approaches the same field value as the physical vev ($\expval{\varphi_2} = v_2$), making a two-step transition possible. Note that similar statements apply to the minimum on the $\varphi_1$ axis.

%%%%%%%%%%%%%%%%%%%%%%%%%%%%%%%%%%%%%%%%%%%%%%%%%%%%%%%%%%
\section{Phase transitions: semianalytical approach}\label{sec:phase_analytical}
%%%%%%%%%%%%%%%%%%%%%%%%%%%%%%%%%%%%%%%%%%%%%%%%%%%%%%%%%%
%
To gain a deeper understanding of the dynamics of two-step FOPTs, we apply the semianalytical results for the Euclidean action obtained in Ref.~\cite{Levi:2022bzt} for scenarios with strong supercooling. Because these results are derived within the single-field approximation, they can be applied only if one of the field directions is fixed.

The finite-temperature effective potential includes thermal corrections via the one-loop thermal functions~\cite{Quiros:1999jp} and contributions from Daisy diagrams~\cite{Arnold:1992rz}. We employ the high-temperature expansion to express the thermal functions in polynomial form, and following Ref.~\cite{Pascoli:2026tuu}, we expand the Daisy potential in a polynomial basis. Then, the thermal effective potential can be written in the polynomial form~\cite{Levi:2022bzt,Pascoli:2026tuu},
\begin{equation}\label{eq:poly_form}
    V_{\rm eff}(\varphi,T) = V_0(T) + \frac{m^2(T)}{2}\varphi^2 - \frac{K(T)}{3}\varphi^3 - \frac{\lambda(T)}{4}\varphi^4\,,
\end{equation}
where each temperature-dependent coefficient is obtained from the polynomial expansion of the thermal and zero-temperature potentials. For illustration we focus on TM1 below; the corresponding expressions for TM2 and TM3 are provided in Appendix~\ref{sec:poly_exp}. 

We are interested in the $(0,0)\rightarrow (v_1,0)/(0,v_2)$ and $(v_1,0)/(0,v_2)\rightarrow (v_1,v_2)$ transitions schematically represented in Fig.~\ref{fig:phase_twostep}. The thermal potential in the $(0,0)\rightarrow (v_1,0)$ direction can be expanded as
%\begin{equation}\label{eq:polyA1_zero}
\begin{align}
    & V_0(T) = -\left(\frac{\pi^2}{15}+ \frac{C_b^{3/2} g_1^3}{12 \pi }\right)T^4 \,, \label{eq:V0_T}\\
    & m^2(T) = \left(\frac{1}{4} -\frac{g_1 C_b^{3/2} c_{2}}{6 \pi }\right)g_1^2 T^2\,,\label{eq:m2T}\\
%    & K(T) = \frac{3 g_1^3 T}{4 \pi} + \frac{g_1^3 T}{4 \pi }\left(C_b^{3/2} c_3 - 1\right) \,,\label{eq:KT}\\
    & K(T) = \Big(C_b^{3/2}c_3 + 2\Big) \frac{ g_1^3 T}{4 \pi}  \,,\label{eq:KT}\\
    & \lambda(T) = -\lambda_1 - \frac{g_1^4}{8\pi^2}\Big[1 -3 \gamma_E+\ln64 + 3\ln(\pi T/\mu)\Big] \nonumber\\
    &\hspace{3.5em} + \frac{g_1^3 C_b^{3/2} c_4}{3 \pi }\,,\label{eq:lambT}
\end{align}
where $C_b = 3/2$  and the constants $c_i$ appear after the Daisy potential is put into polynomial form by a Gram-Schmidt decomposition. The expressions for these coefficients can be found in Ref.~\cite{Pascoli:2026tuu}, and are evaluated at $\phi_{1,2}/T = 5$ because to analyze the tunneling dynamics it is sufficient to consider the effective potential up to this value. Identical expressions apply for $(0,0)\rightarrow (0,v_2)$ with the replacements, $g_1 \rightarrow g_2$ and $\lambda_1 \rightarrow \lambda_2$ in Eqs.~\eqref{eq:V0_T}--\eqref{eq:lambT}. The potential for the $(v_1,0)\rightarrow (v_1,v_2)$ transition is
\begin{align}
    & V_0(T) = \frac{g_1^4 v_1^4}{32 \pi ^2} [1-3 \gamma_E + 3\ln(4\pi T/\mu)] - \frac{g_1^3  v_1^3}{4 \pi } T \nonumber \\
    &\hspace{4em} + \frac{1}{8} \left(g_1^2 v_1^2 T^2+2 \lambda_1 v_1^4\right) \nonumber\\
    &\hspace{4em} -\left(\frac{\pi^2}{15}+ \frac{C_b^{3/2} g_2^3}{12 \pi }\right)T^4 + V_{\rm Daisy}(v_1,0)\,,\label{eq:V0_T_newdir}\\
    & m^2(T) = \frac{1}{4} \left(g_2^2 T^2+2 \lambda_{12} v_1^2\right) -\frac{g_2^3 C_b^{3/2} c_{2}}{6 \pi }T^2 \,, \label{eq:m2T_newdir}\\
    & K(T) = \Big(C_b^{3/2}c_3 + 2\Big) \frac{ g_2^3 T}{4 \pi} \,, \label{eq:KT_newdir}\\
    & \lambda(T) = -\lambda_2 - \frac{g_2^4}{8\pi^2}\Big[1 -3 \gamma_E+ 6\ln2 + 3\ln(\pi T/\mu)\Big] \nonumber\\
    &\hspace{3.5em} + \frac{g_2^3 C_b^{3/2} c_4}{3 \pi } \label{eq:lambT_newdir}\,,
\end{align}
where 
\begin{equation}\label{eq:vdaisy_dir}
    \begin{aligned}
    &V_{\rm Daisy}(v_1,0) = \frac{1}{48\pi}\Big[4 g_1^3 v_1^3 T-3 \sqrt{6} g_2^3 T^4\Big] \\
    &\hspace{6.5em} - \frac{T}{24\sqrt{2}\pi}\Big[g_1^2 \left(3 T^2+2 v_1^2\right)\Big]^{3/2}\,.
    \end{aligned}
\end{equation}
The potential for the $(0,v_2)\rightarrow (v_1,v_2)$ transition is obtained from Eqs.~\eqref{eq:V0_T_newdir}--\eqref{eq:vdaisy_dir} by making the replacements $g_1 \leftrightarrow g_2$, $v_1 \rightarrow v_2$ and $\lambda_1 \rightarrow \lambda_2$. With the potential in the polynomial form of Eq.~\eqref{eq:poly_form}, we employ the results in Ref.~\cite{Levi:2022bzt} to evaluate the Euclidean action. Defining $\kappa(T) \equiv -\lambda(T)m^2(T)/K(T)^2$, the 3-dimensional action of can be written as~\cite{Levi:2022bzt}
\begin{equation}\label{eq:action}
    S_3(T) \simeq \begin{cases}\dfrac{m^3(T)}{K^2(T)} \dfrac{2 \pi}{3\left(\kappa-\kappa_c\right)^2} \bar{B}_3(\kappa)\,, & \kappa>0 \\ \dfrac{m^3(T)}{K^2(T)} \dfrac{27 \pi}{2}\left(\dfrac{1+e^{-1 / \sqrt{|\kappa|}}}{1+\frac{9}{2}|\kappa|}\right)\,, & \kappa<0\,,\end{cases}
\end{equation}
where $\kappa_c = \kappa(T=T_c) = 2/9$ with $T_c$ the critical temperature, and~\cite{Levi:2022bzt} 
\begin{equation}
 \begin{aligned}
    \bar{B}_3(\kappa)=\frac{16}{243}\bigg[1-38.23\left(\kappa-\frac{2}{9}\right) + 115.26\left(\kappa-\frac{2}{9}\right)^2  \\
   + 58.07\sqrt{\kappa}\left(\kappa-\frac{2}{9}\right)^2+229.07\kappa\left(\kappa-\frac{2}{9}\right)^2\bigg]\,.\nonumber
     \end{aligned}
\end{equation}

%%%%%%%%%%%%%%%%%%%%%%%%%%%%%%%%%%%%%%%%%%%%%%%%%%%%%%%%%%
%\textit{ \textbf{Gravitational waves}.}
\section{Gravitational waves}\label{sec:GWs}
%%%%%%%%%%%%%%%%%%%%%%%%%%%%%%%%%%%%%%%%%%%%%%%%%%%%%%%%%%

The GW spectrum is fully characterized by four thermodynamic parameters: 1) percolation temperature, $T_p$, defined as the temperature at which the true vacuum occupies 34\% of the volume of the Universe. Since we consider two-step transitions following the pattern $(0,0) \rightarrow (v_1,0)/(0,v_2) \rightarrow (v_1,v_2)$, we require that $T_p[(0,0) \rightarrow (v_1,0)/(0,v_2)] > T_p[(v_1,0)/(0,v_2) \rightarrow (v_1,v_2)]$ to ensure that the first transition occurs before the second one begins. 
We additionally require both FOPTs to complete by checking that the false-vacuum fraction is decreasing at percolation;
2) the reheating temperature after the completion of the phase transition, $T_\mathrm{RH} \approx T_p(1+\alpha)^{1/4}$; 3) the transition strength $\alpha$ defined as the 
ratio of the total latent heat to the total radiation energy density at percolation; and 4)
the inverse time duration of the FOPT normalized to the Hubble rate at percolation $\beta/H(T_p)$. Since both steps of the FOPT exhibit mild to strong supercooling, we approximate the bubble wall velocity to $v_w = 1$. 

The two-peak GW spectrum is usually computed by evaluating a single-step template~\cite{Caprini:2024hue} (which relies on the above parameters) for each step and summing them. The template implicitly assumes (i)~that the Universe at production contains only radiation plus the vacuum energy released in that step, and (ii)~that the expansion from production until today is adiabatic. In a two-step transition, the first step can violate both assumptions. 
To ensure that the vacuum energy that will be released by the second step is negligible compared to the total energy density during the first step, we require $T_{\rm RH}[(0,0) \rightarrow (v_1,0)/(0,v_2)] > T_{\rm RH}[(v_1,0)/(0,v_2) \rightarrow (v_1,v_2)]$. 
 The latent heat released by the second step is an entropy injection that dilutes the GWs from the first step, and shifts the spectrum to lower frequency and lower amplitude $\Omega_1$:
\begin{equation}\label{eq:shift_rel}
   f \rightarrow {f\over (1+\alpha_2)^{1/4}} \,, \quad \Omega_1(f)\to \frac{1}{1+\alpha_2}\,\Omega_1\!\left((1+\alpha_2)^{1/4}f\right)\,,
\end{equation}
where $\alpha_2$ is the strength of the second FOPT. GWs from the second step are not affected. 

To construct the two-peak spectrum, we account for the entropy dilution of the GW signal from the first FOPT, and then sum the contributions from each phase transition.  

%%%%%%%%%%%%%%%%%%%%%%%%%%%%%%%%%%%%%%%%%%%%%%%%%%%%%%%%%
%\textit{\textbf){Numerical results}.}
\section{Numerical results}\label{sec:num_results}
%%%%%%%%%%%%%%%%%%%%%%%%%%%%%%%%%%%%%%%%%%%%%%%%%%%%%%%%%
\begin{table}[t]
\centering
\begin{tabular}{lll}
\toprule
Parameter & Sampling & Range \\
\midrule
$g_1$                 & Uniform & $[0.1,\, 1]$ \\
$g_2$                 & Uniform & $[0.1,\, 1]$ \\
$v_1\ (\mathrm{GeV})$ & Log     & $[150,\, 10^{10}]$ \\
$v_2\ (\mathrm{GeV})$ & Log     & $[150,\,10^{10}]$ \\
$|\lambda_{12}|$        & Log     & $[10^{-10}, 10^{-2}]$ \\
$g_3$                 & Log     & $[10^{-5},\,1]$ \\
$y$                   & Log     & $[10^{-5},\,1]$ \\
\bottomrule
\end{tabular}
\caption{Sampling methodology and input parameter ranges. }
\label{tab:scan_ranges}
\end{table}
\begin{table}[t]
    \centering
    \begin{tabular}{llll}
        \hline
        {} & BP-TM1 & BP-TM2 & BP-TM3 \\
        \hline
        $g_1$ & $0.81$ & $0.83$ & $0.90$ \\[0.1em]
        $g_2$ & $0.64$ & $0.86$ & $0.60$ \\[0.1em]
        $\lambda_{12}$ & $-2.76\times 10^{-5}$ & $-2.12\times 10^{-9}$ & $-1.90\times 10^{-8}$ \\[0.1em]
        $v_1~(\mathrm{GeV})$ & $2.25\times 10^8$ & $18395$ & $7.22\times 10^5$ \\[0.1em]
        $v_2~(\mathrm{GeV})$ & $1.84\times 10^8$ & $284.61$ & $6.02\times 10^7$ \\[0.1em]
        $g_3$ & $--$ & $0.0095$ & $--$ \\[0.1em]
        $y$ & $--$ & $--$ & $0.0031$ \\
        \hline
        $\alpha_1$ & $89.93$ & $1.76$ & $7.20\times 10^7$ \\
        $\alpha_2$ & $51.33$ & $0.49$ & $16.76$ \\
        $[\beta/H(T_p)]_1$ & $27.19$ & $76.56$ & $33.90$ \\
        $[\beta/H(T_p)]_2$ & $254.22$ & $107.99$ & $15.79$ \\
        $[T_c]_1~(\mathrm{GeV})$ & $6.58\times 10^7$ & $5493.6$ & $1.21\times 10^7$ \\
        $[T_p]_1~(\mathrm{GeV})$ & $5.41\times 10^6$ & $1201.4$ & $3.56\times 10^4$ \\
        $[T_{\rm RH}]_1~(\mathrm{GeV})$ & $1.67\times 10^7$ & $1548.2$ & $3.28\times 10^6$ \\
        $[T_c]_2~(\mathrm{GeV})$ & $3.98\times 10^7$ & $88.99$ & $2.40\times 10^5$ \\
        $[T_p]_2~(\mathrm{GeV})$ & $3.98\times 10^6$ & $26.45$ & $2.91\times 10^4$ \\
        $[T_{\rm RH}]_2~(\mathrm{GeV})$ & $1.07\times 10^7$ & $29.24$ & $5.98\times 10^4$ \\
        $\mathrm{SNR_{LISA}}$ & $1.15\times 10^{-5}$ & $674.7$ & $3400$ \\
        $\mathrm{SNR_{ET}}$ & $3108$ & $1.02\times 10^{-5}$ & $1479$ \\
        \hline
    \end{tabular}%
    \caption{Benchmark points for each toy model. The SNRs correspond to an observation time of 4 years.}
    \label{tab:benchmarks}
\end{table}

\begin{figure}
    \centering
    \includegraphics[width=0.5\textwidth]{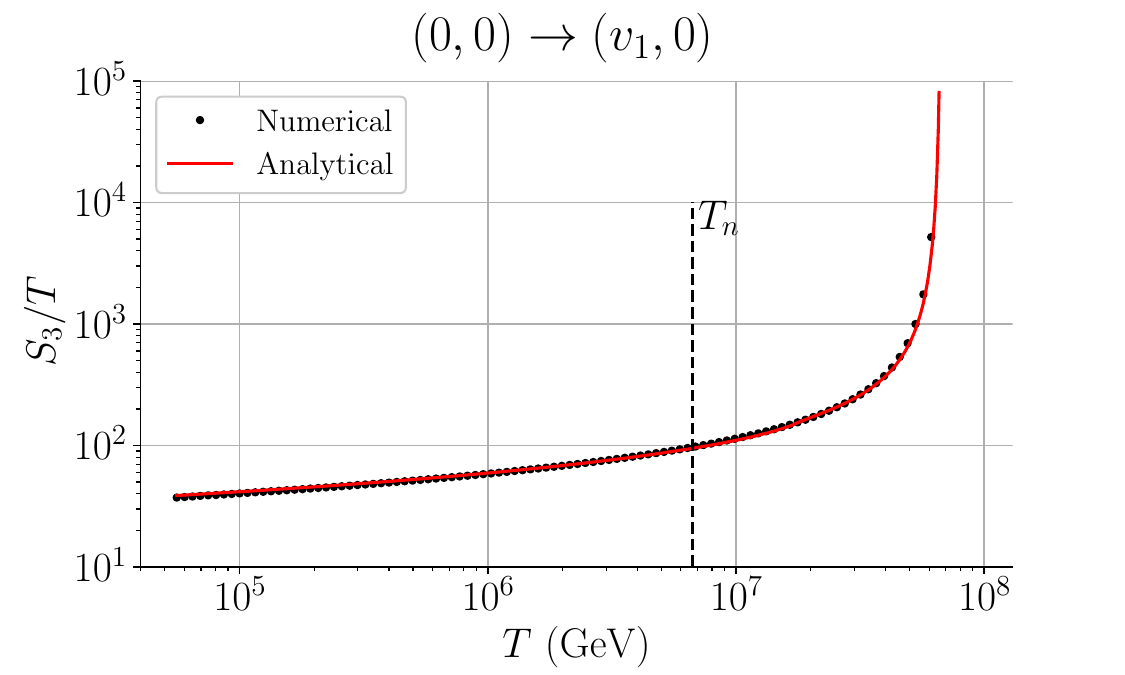}
    \includegraphics[width=0.5\textwidth]{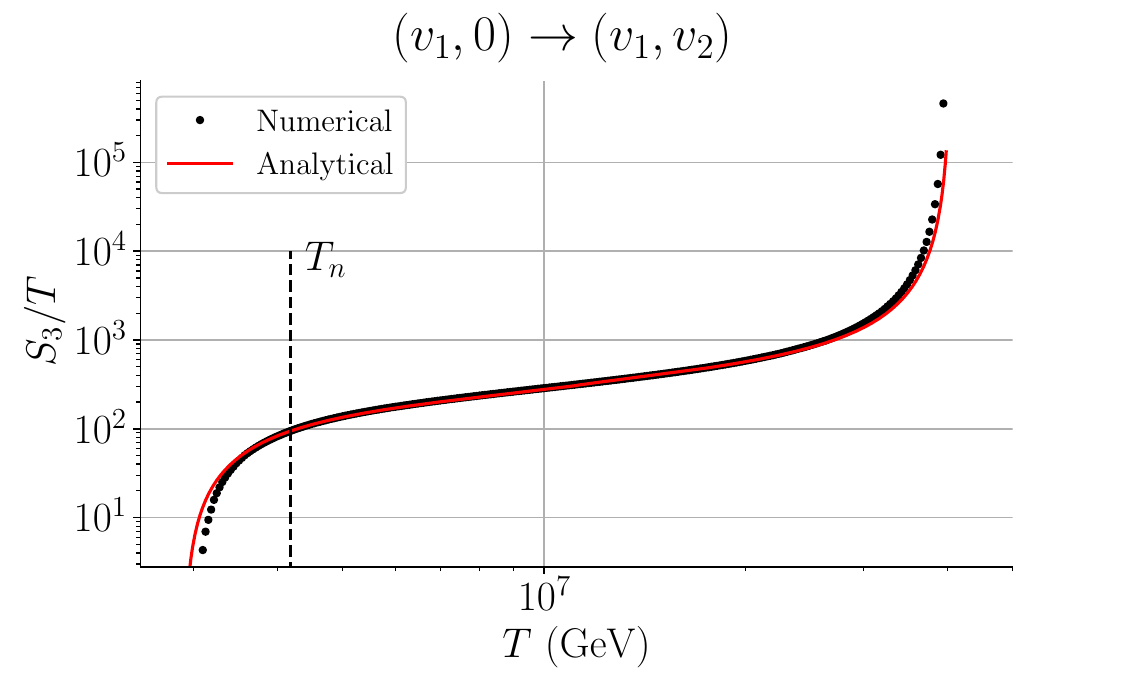}
    \caption{Comparison between the semianalytical and numerical calculations of the Euclidean action for BP-TM1. The top (bottom) panel shows the comparison for the $(0,0)\rightarrow (v_1,0)$ ($(v_1,0)\rightarrow (v_1,v_2)$) transition. The nucleation temperature $T_n$ is marked for each FOPT. }
    \label{fig:action_comparison}
\end{figure}

We perform a scan over the model parameters. For TM1, these are $g_1$, $g_2$, $v_1$, $v_2$ and $\lambda_{12}$. TM2 has an additional gauge coupling $g_3$, and TM3 has an additional Yukawa coupling $y$; see Eq.~\eqref{eq:one1loop}. The scan ranges for each parameter are as in Table~\ref{tab:scan_ranges}. 
We identify a benchmark point for each toy model in Table~\ref{tab:benchmarks}. The model parameters and corresponding thermodynamic and GW observables are listed. The thermodynamic parameters are computed using the tunneling action in Eq.~\eqref{eq:action}. We checked our results obtained from Eq.~\eqref{eq:action} against a numerical implementation in \texttt{CosmoTransitions}~\cite{Wainwright:2011kj} and found subpercent agreement near the nucleation temperature $T_n$ (see Fig.~\ref{fig:action_comparison}).  We further compared our results with a  \texttt{CosmoTransitions} implementation using the full thermal potential (without relying on the polynomial expansion). For field regions relevant for tunneling, $\varphi/T \lesssim 5$, we found agreement within 1\% for $(0,0)\rightarrow (v_1,0)/(0,v_2)$ transitions, and an agreement within 10\% for $(v_1,0)/(0,v_2) \rightarrow (v_1,v_2)$ transitions.

From Table~\ref{tab:benchmarks}, note that both FOPTs undergo at least mild supercooling, with $\alpha_1$ and $\alpha_2$ above $\sim 0.5$ and $\beta/H(T_p)$ below $\sim 250$ for all three benchmarks.

\begin{figure}[t]
    \centering
    \includegraphics[width=0.5\textwidth]{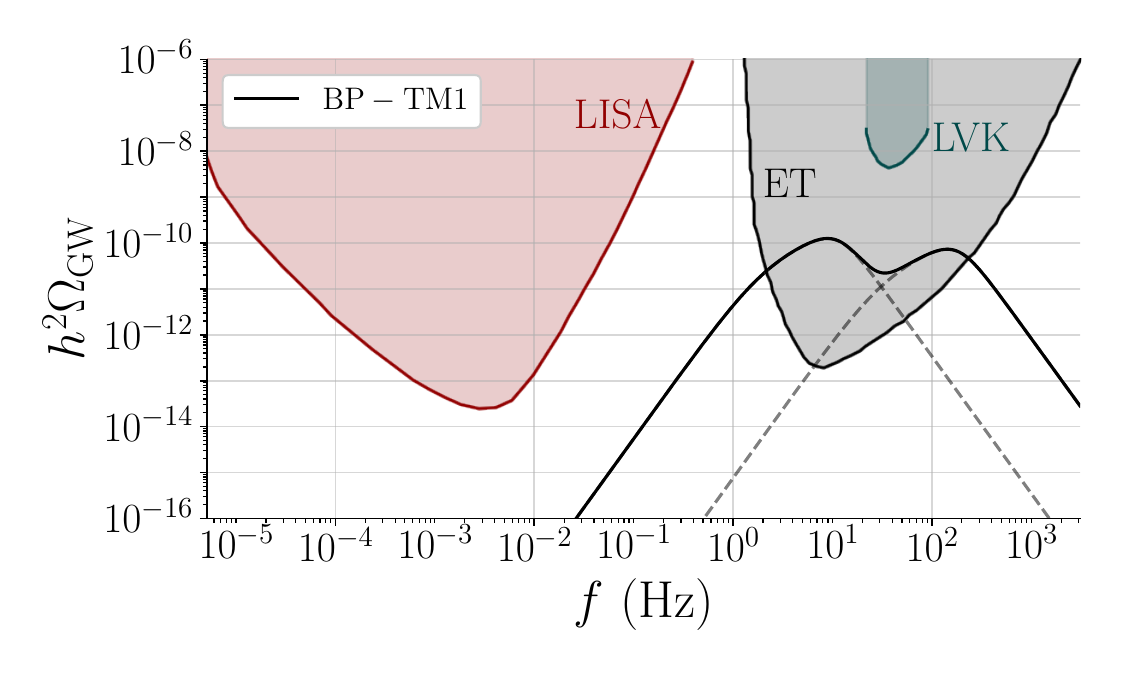} \\
    \includegraphics[width=0.5\textwidth]{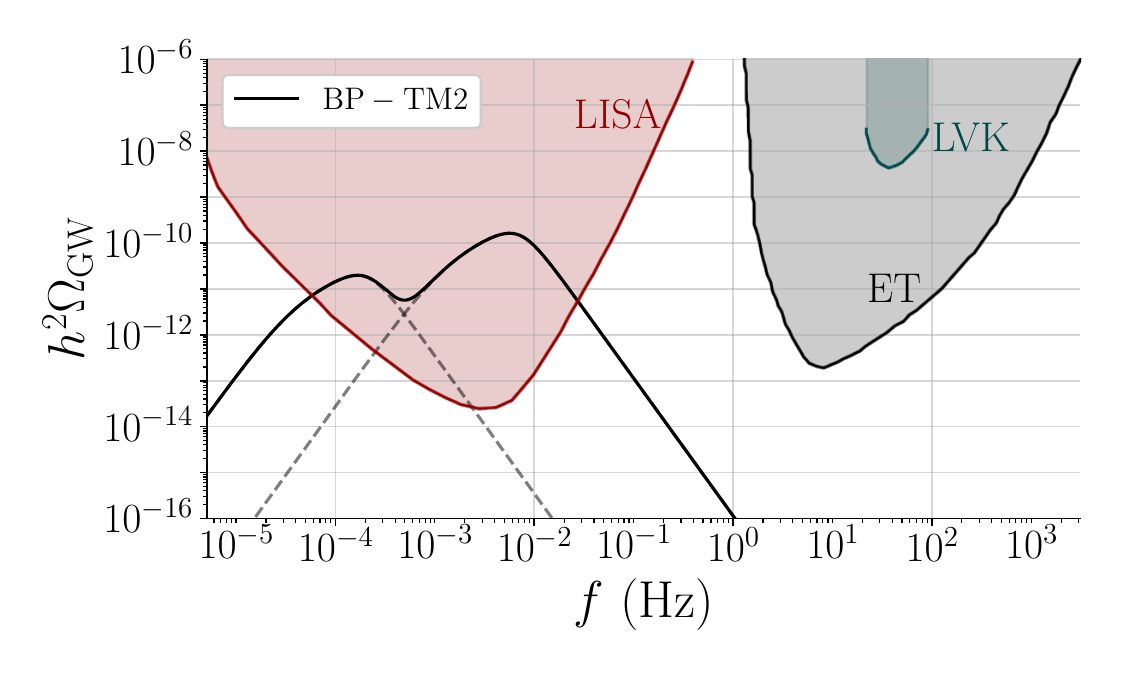} \\
    \includegraphics[width=0.5\textwidth]{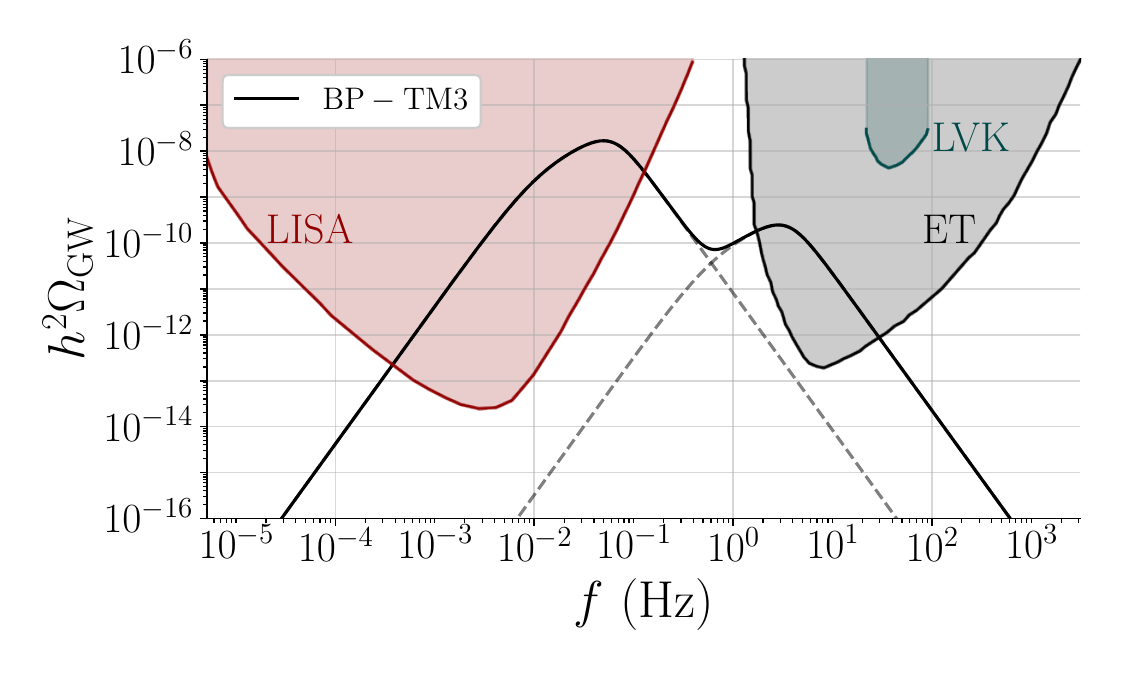}
    \caption{Two-peak GW spectrum for the three benchmark points in Table~\ref{tab:scan_ranges}. The spectrum from each FOPT is shown as a dashed curve, and their sum is the solid curve. We also show the current $2\sigma$ LIGO-Virgo-KAGRA (LVK) bound~\cite{Virgo:2025aai} and the sensitivity curves for LISA~\cite{LISA:2017pwj} and ET~\cite{Punturo:2010zz}. }
    \label{fig:spectra_TM}
\end{figure}

As shown in Fig.~\ref{fig:spectra_TM}, the three benchmark points exhibit a two-peak spectrum. For BP-TM1 and BP-TM2, both peaks lie within reach of the Einstein Telescope (ET) and of LISA, respectively, and for BP-TM3, one of the peaks lies in the LISA band and the other peak is within ET's band. The large separation in the two peak frequencies is correlated with a large difference in the reheating temperatures (which differ by more than an order of magnitude for BP-TM2). Furthermore, the strongly supercooled nature of both transitions yields large signal-to-noise ratios (SNRs). In particular, for BP-TM3, both LISA and ET have an SNR above 1000.

\begin{figure*}[t]
    \centering
    \includegraphics[width=1.0\textwidth]{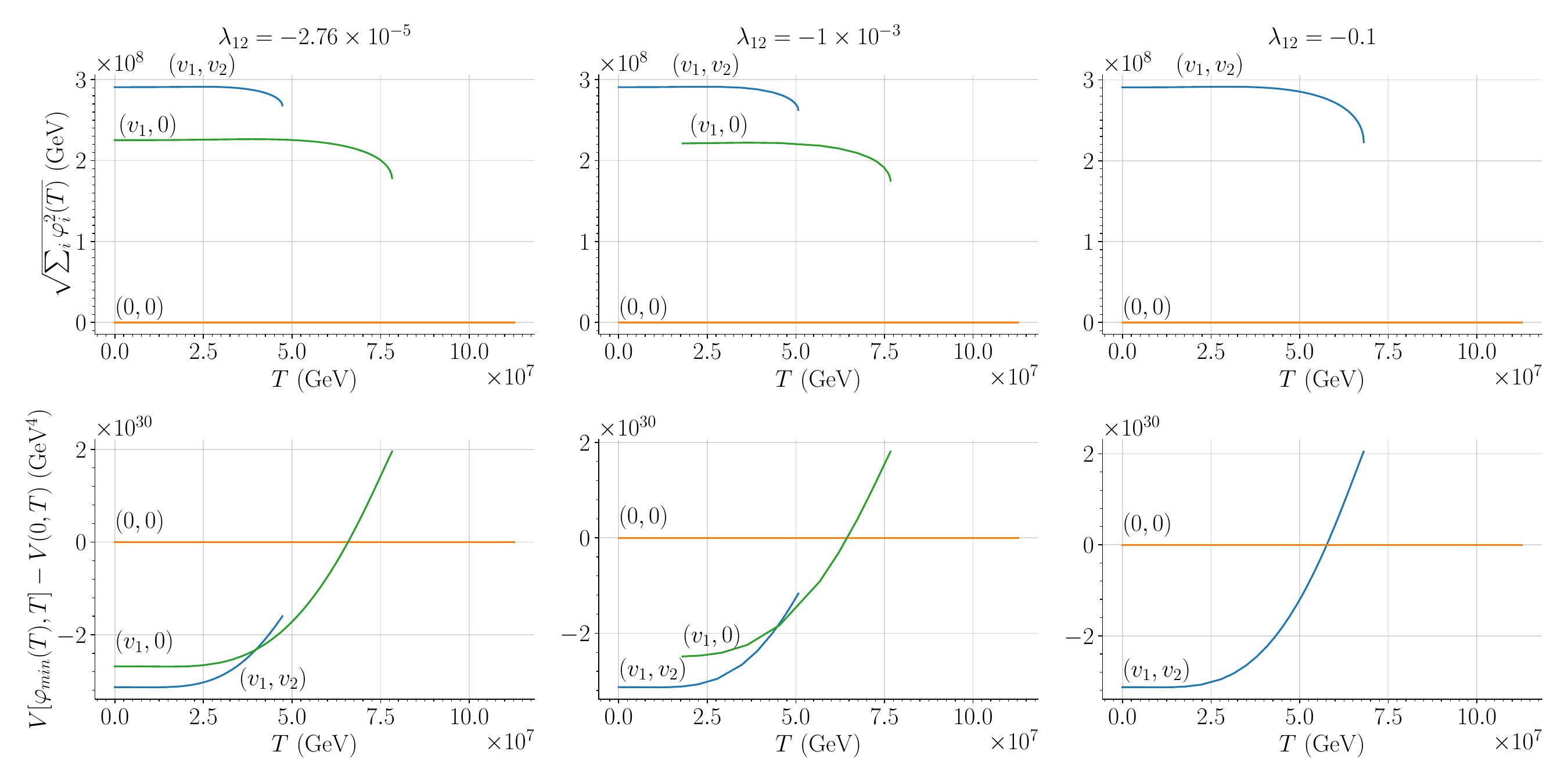}
    \caption{Temperature evolution of the phases for BP-TM1 for three values of $\lambda_{12}$; all other parameters are fixed to the values in Table~\ref{tab:benchmarks}. The top row shows the square root of the squared sum of the phases, and the bottom row shows the potential difference between each minimum and the origin. }
    \label{fig:phase_evolution}
\end{figure*}
\begin{figure*}[t]
    \centering
    \includegraphics[width=0.34\textwidth]{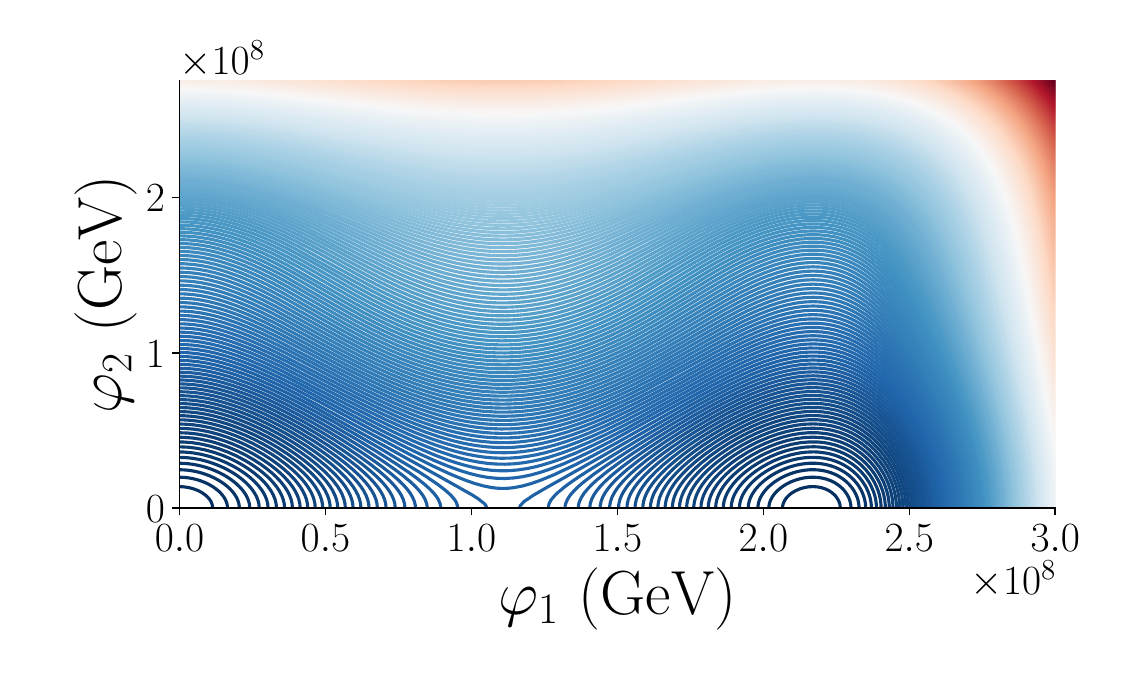}
    \hspace*{-1.4em}
    \includegraphics[width=0.34\textwidth]{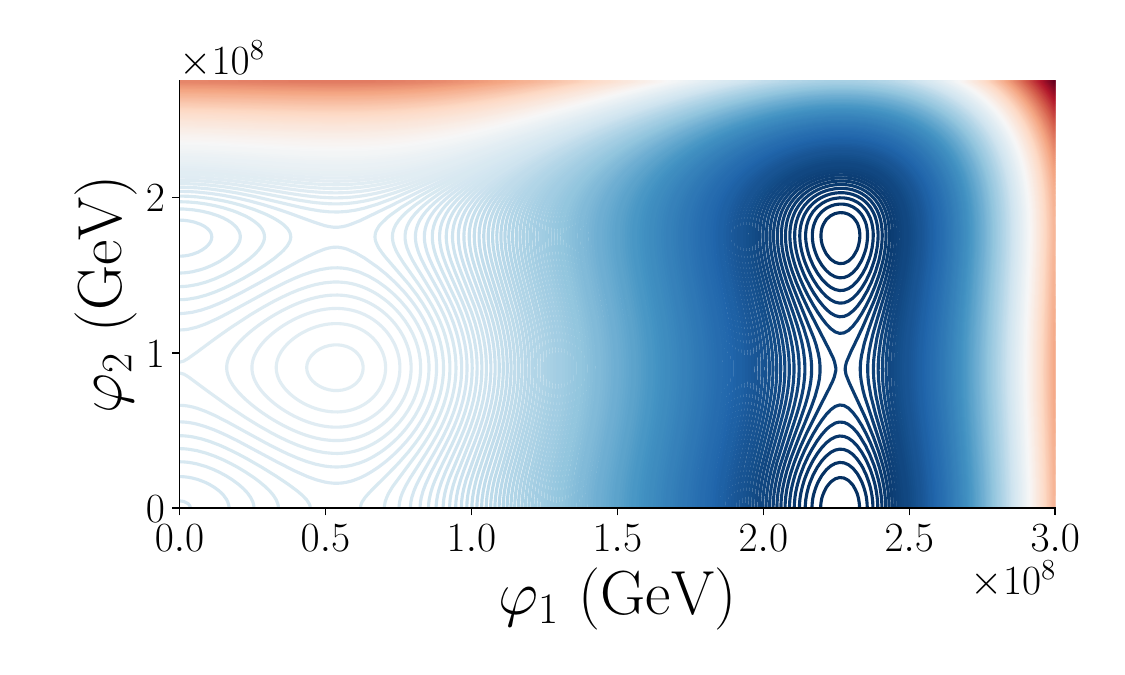}
    \hspace*{-1.4em}
    \includegraphics[width=0.34\textwidth]{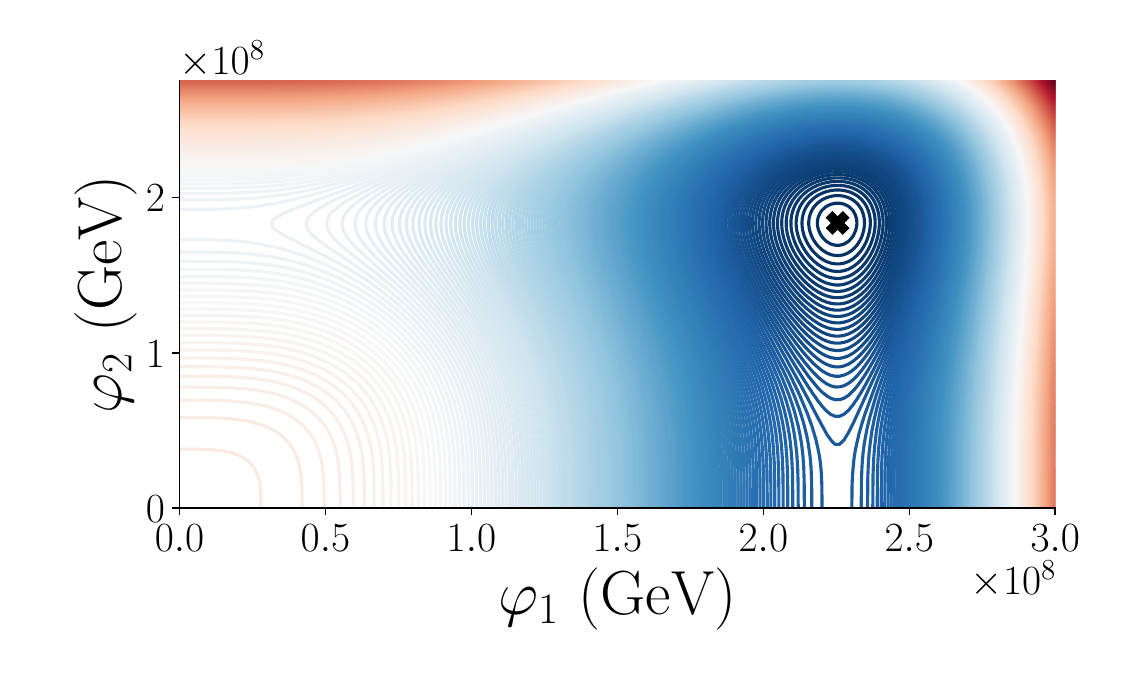}
    \caption{Iso-$V_{\rm eff}$ contours for BP-TM1 at $T = [T_c]_1$ (left panel), $T = [T_c]_2$ (middle panel) and $T = 0$ (right panel). The global minimum $(v_1,v_2)$ at $T=0$ is marked in the right panel. }  \label{fig:potential_twostep}
\end{figure*}

\begin{figure*}[t]
    \centering
    \includegraphics[width=0.495\textwidth]{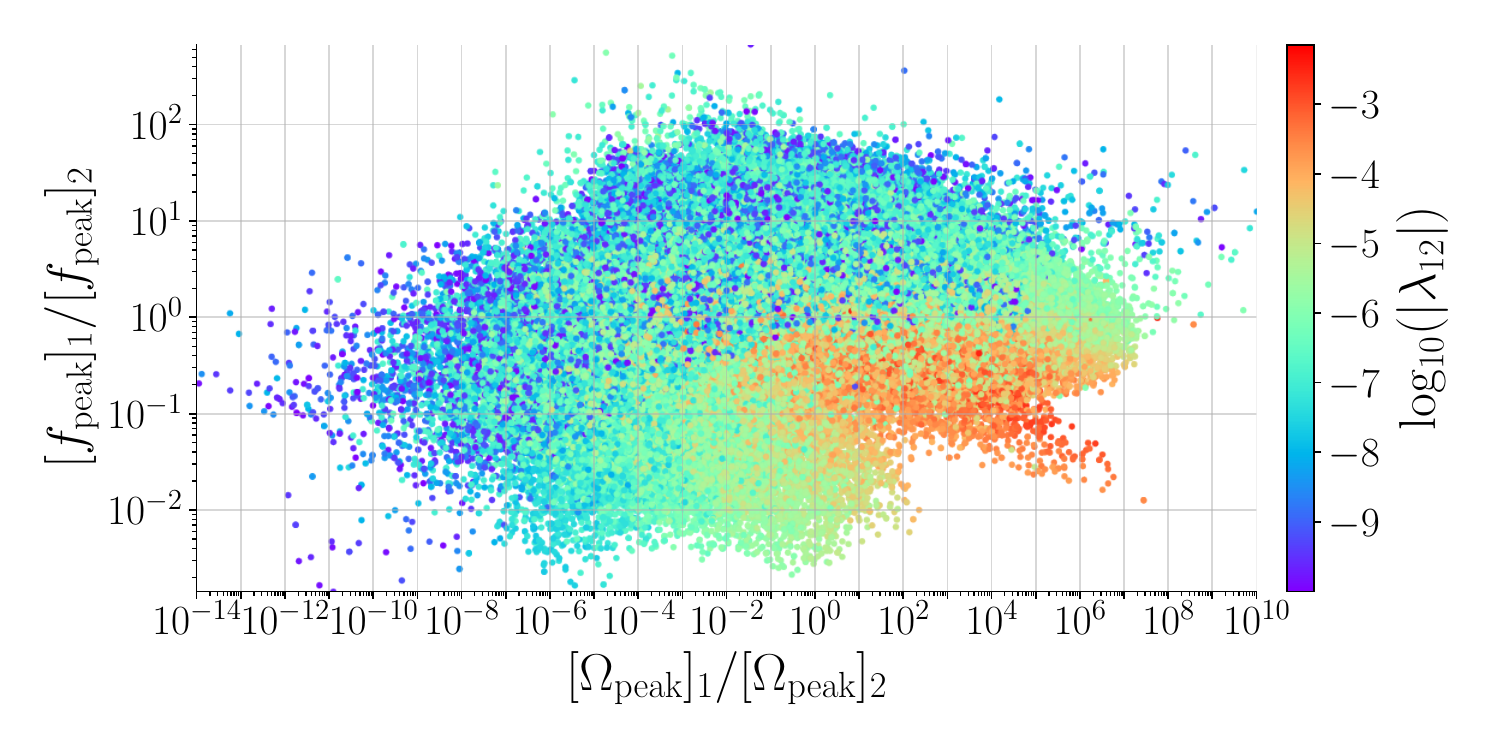}
    \includegraphics[width=0.495\textwidth]{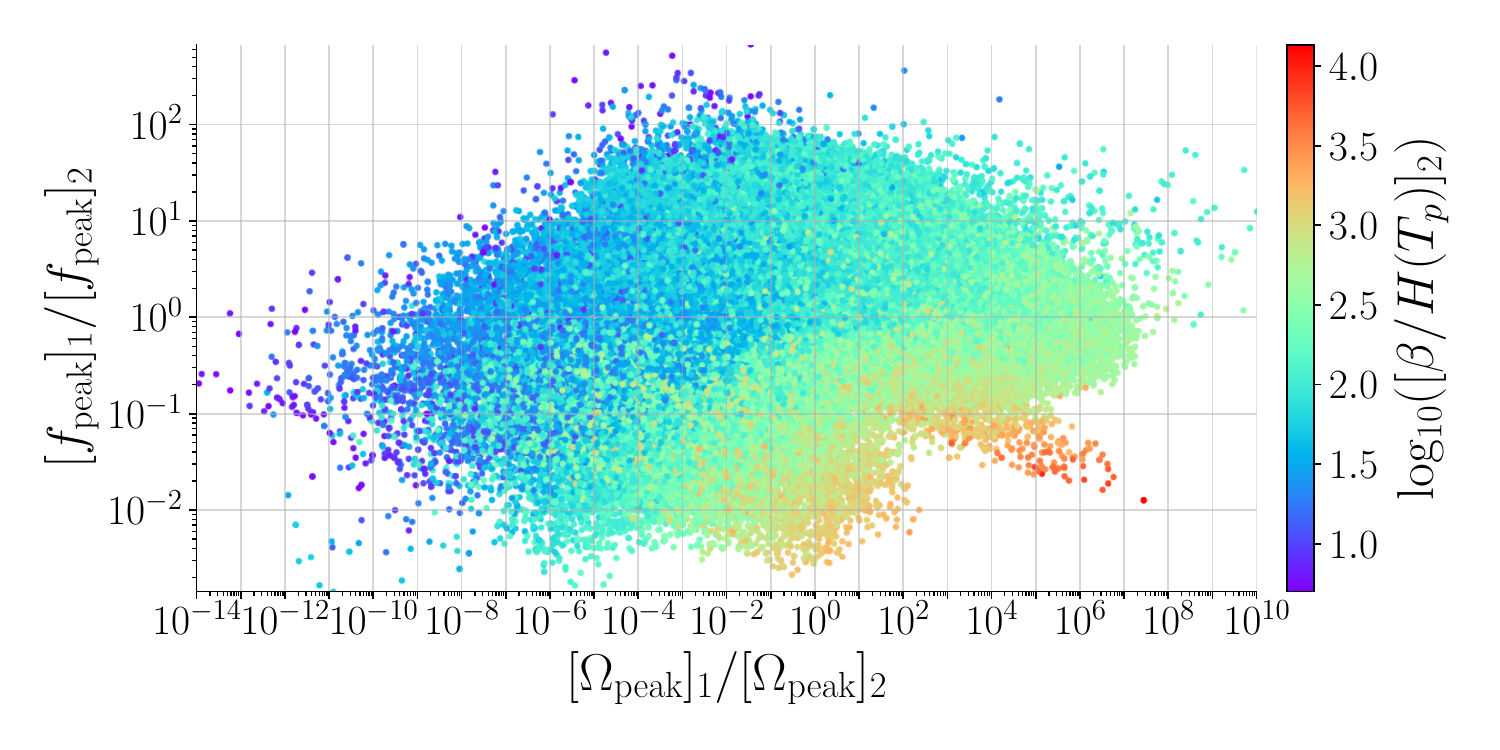}
    \includegraphics[width=0.495\textwidth]{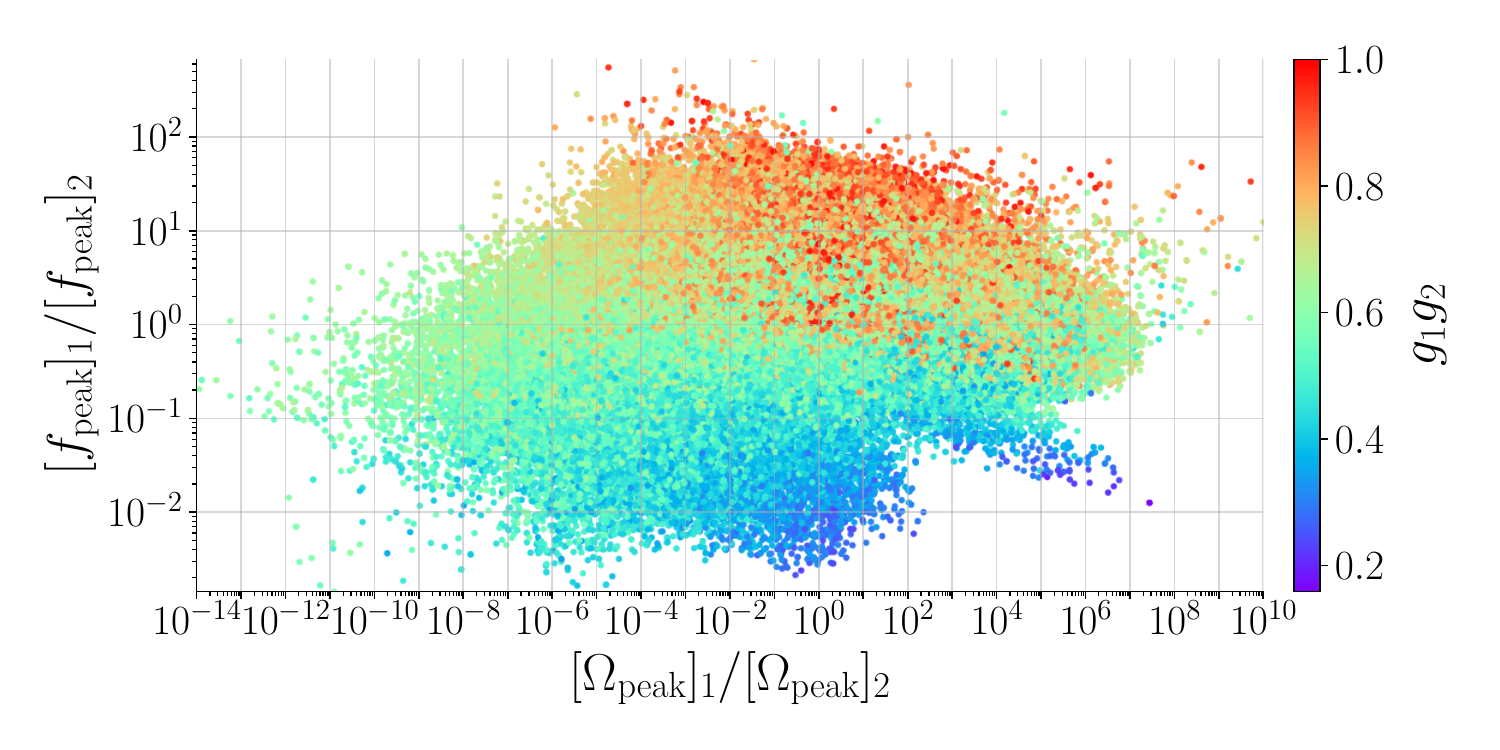}
    \includegraphics[width=0.495\textwidth]{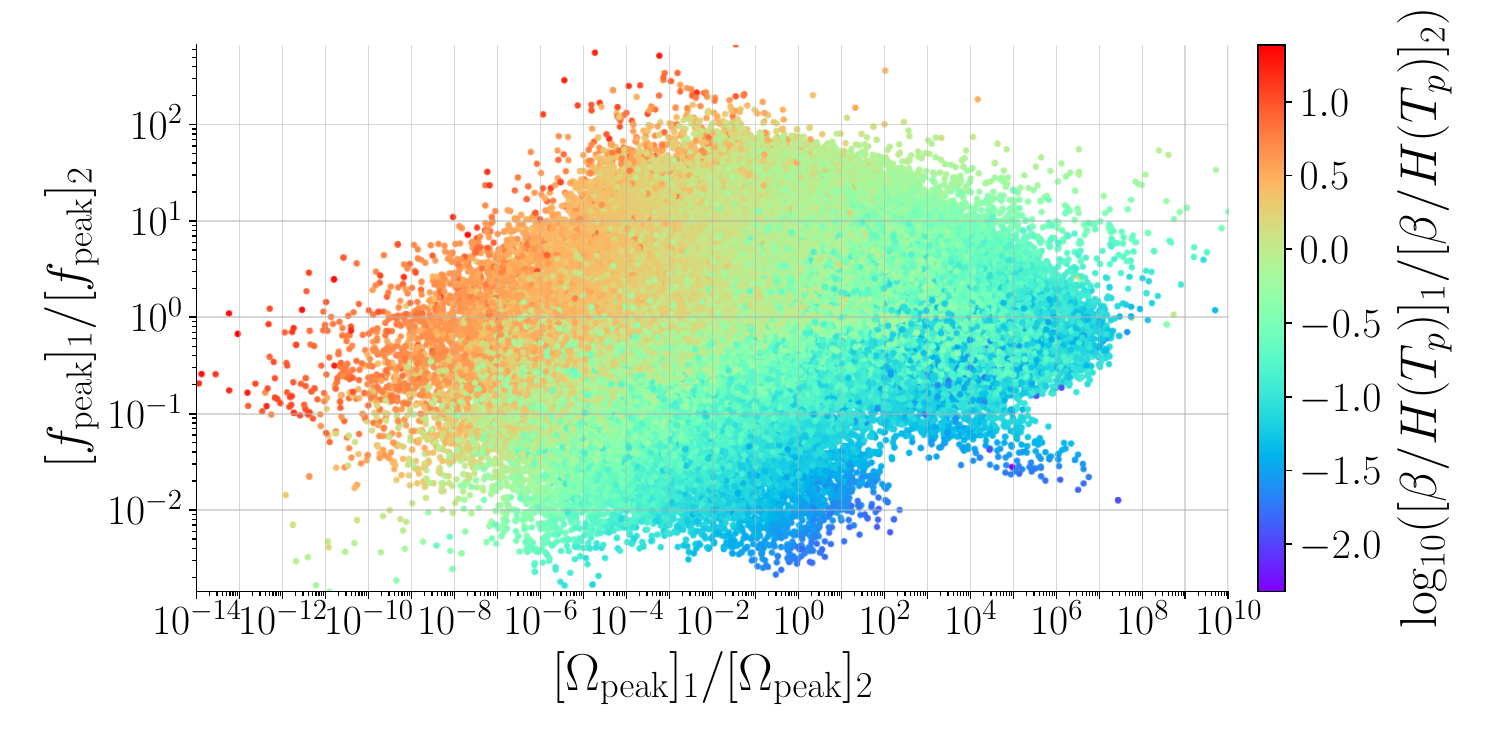}
    \caption{Scatter plots for TM1, of the ratio of the peak frequencies as a function of the ratio of the corresponding peak amplitudes of the two-peak GW spectrum. The color bar indicates $\mathrm{log_{10}}(|\lambda_{12}|)$ in the top-left panel, $g_1g_2$ in the bottom-left panel, $\mathrm{log_{10}}([\beta/H(T_p)]_2)$ in the top-right panel and $\mathrm{log_{10}}([\beta/H(T_p)]_1/[\beta/H(T_p)]_2))$ in the bottom-right panel. }
    \label{fig:scan_TM1}
\end{figure*}

To better understand the dynamics of a two-step phase transition, in Fig.~\ref{fig:phase_evolution} we show the thermal evolution of the phases for BP-TM3 in Table~\ref{tab:benchmarks} for three values of $\lambda_{12}$. We complement this with Fig.~\ref{fig:potential_twostep}, in which we show contour plots of the thermal effective potential at three important temperatures for the value of $\lambda_{12}$ in the left panel of Fig.~\ref{fig:phase_evolution}. From the 
left panels of Fig.~\ref{fig:phase_evolution} we observe that at high temperatures, the global minimum of the potential is at $(\varphi_1,\varphi_2) = (0,0)$, as expected. As the Universe cools, a new local minimum at $(\varphi_1,\varphi_2) = (v_1,0)$ appears at  $T \approx 7.82\times 10^7~\mathrm{GeV}$ (in the upper-left panel). The $(0,0) \rightarrow (v_1,0)$ transition becomes possible once the temperature drops below the critical temperature $[T_c]_1 \approx 6.58\times 10^7~\mathrm{GeV}$, where $V(v_1,0,[T_c]_1) = V(0,0,[T_c]_1)$ (see the bottom-left panel, and the left panel of Fig.~\ref{fig:potential_twostep}). As the temperature falls further, a second minimum at $(v_1,v_2)$ emerges at $T \approx 4.73\times 10^7~\mathrm{GeV}$ (in the upper-left panel).  The second transition $(v_1,0) \rightarrow (v_1,v_2)$ becomes possible once the temperature drops below the critical temperature at $[T_c]_2 \approx 3.98\times 10^7~\mathrm{GeV}$, where $V(v_1,v_2,[T_c]_2) = V(0,v_2,[T_c]_2)$ (see the middle panel of Fig.~\ref{fig:potential_twostep}). After the second transition, the global minimum is located at $(\varphi_1,\varphi_2) = (v_1,v_2)$, and is marked in the right panel of Fig.~\ref{fig:potential_twostep}.
From the middle and right panels of Fig.~\ref{fig:phase_evolution} we find that as 
$\lambda_{12}$ decreases, the temperature interval over which the intermediate minimum $(0,v_2)$ persists gets shorter. For sufficiently small $\lambda_{12}$, this minimum disappears entirely, and the two-step transition becomes a single-step $(0,0) \rightarrow (v_1,v_2)$ transition. This can be understood from Eq.~\eqref{eq:vev_dir1}, where large negative
$\lambda_{12}$ pushes the minimum to zero. 

In Fig.~\ref{fig:scan_TM1}, for TM1 we show scatter plots of the ratio of the peak frequencies as a function of the ratio of the corresponding peak amplitudes of the two-peak GW spectrum. The subscripts $1,2$ in the axis labels correspond to the transition with the higher/lower percolation temperature. From the top-left panel we see a clear general trend: larger values of $|\lambda_{12}|$ correspond to larger peak amplitude ratios and smaller peak frequency ratios. This can be understood from Fig.~\ref{fig:phase_evolution}. As $|\lambda_{12}|$ increases, the intermediate phase exists over a progressively narrower temperature range, which compresses the temperature interval available for the second transition. Thus, $[\beta/H(T_p)]_{2}$ increases (as confirmed by the top-right panel of Fig.~\ref{fig:scan_TM1}), and shifts the peak frequency to higher values such that $[f_{\rm peak}]_2 > [f_{\rm peak}]_1$. Concomitantly, the larger value of $[\beta/H(T_p)]_{2}$ suppresses the peak amplitude, resulting in $[\Omega_{\rm peak}]_2 < [\Omega_{\rm peak}]_1$.

The bottom-left panel of Fig.~\ref{fig:scan_TM1} shows that the ratio of the peak frequencies generally increases with the product of the gauge couplings, while the ratio of peak amplitudes does not exhibit a strong correlation with it. The former behavior can be understood from a similar scaling in the ratio of the $\beta/H(T_p)$ parameters with the ratio of the peak frequencies in the 
bottom-right panel. This is partially analogous to  $\mathrm{U(1)}^\prime$ conformal models, where larger $\mathrm{U(1)}^\prime$ gauge couplings lead to larger $\beta/H(T_p)$ values~\cite{Goncalves:2024lrk}. 

We find similar trends for TM2 and TM3.
\begin{figure}[t]
    \centering
    \includegraphics[width=0.5\textwidth]{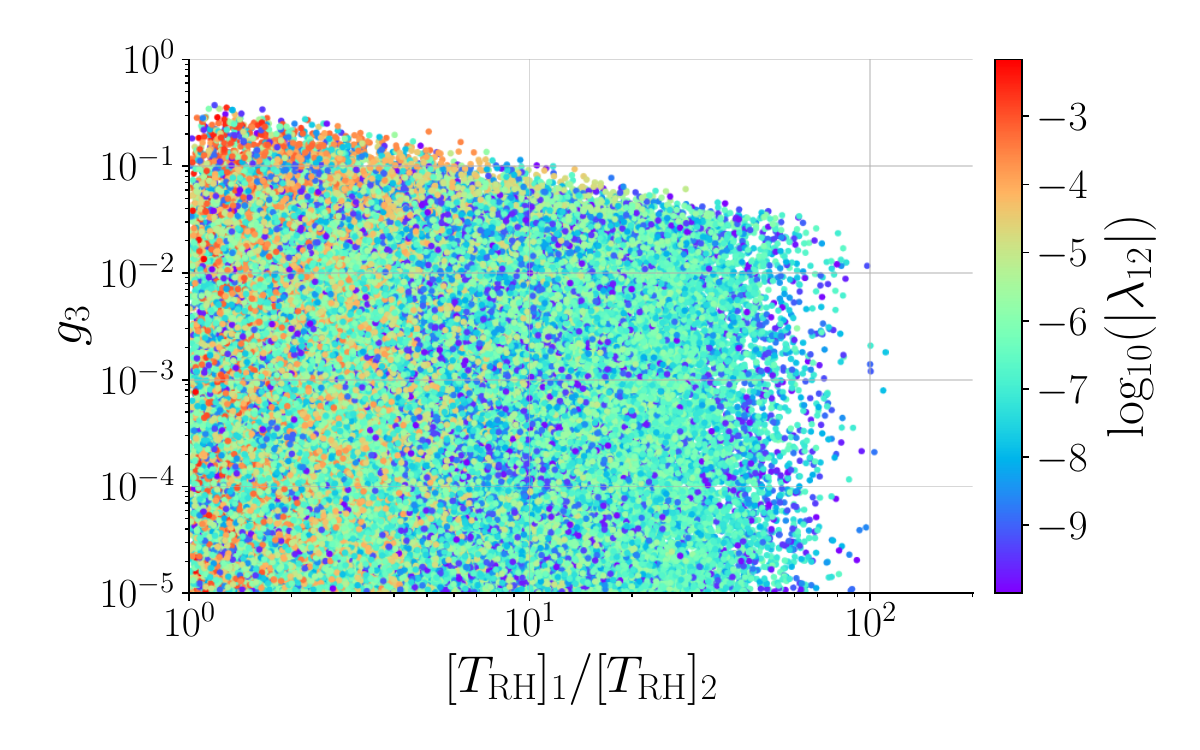} \\
    \includegraphics[width=0.5\textwidth]{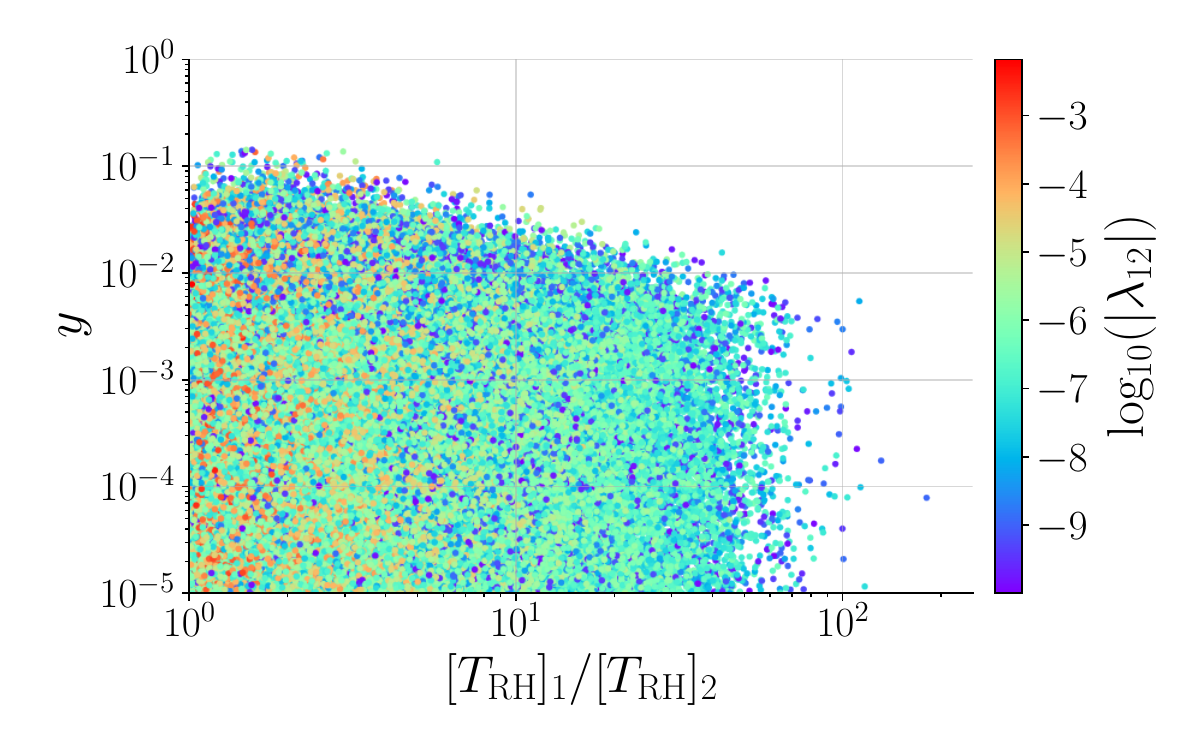} \\
    \caption{Scatter plots of the gauge coupling $g_3$ in TM2 (top panel) and the Yukawa coupling $y$ in TM3 (bottom panel) as a function of the ratio of percolation temperatures, with $\mathrm{log_{10}}(|\lambda_{12}|)$ in the color bar. }
    \label{fig:scatter_y_g3}
\end{figure}
\begin{figure}[t]
    \centering
    \includegraphics[width=0.5\textwidth]{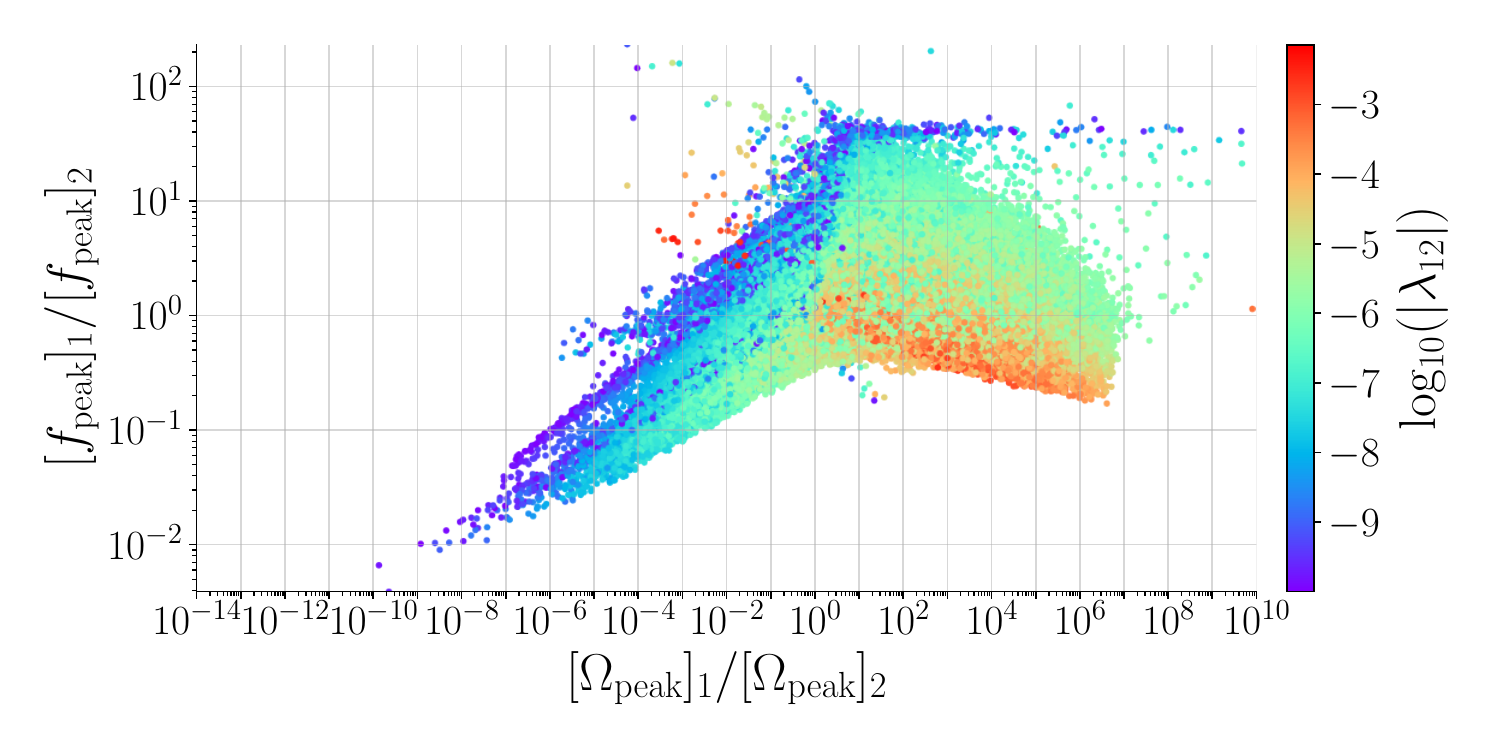} \\
    \includegraphics[width=0.5\textwidth]{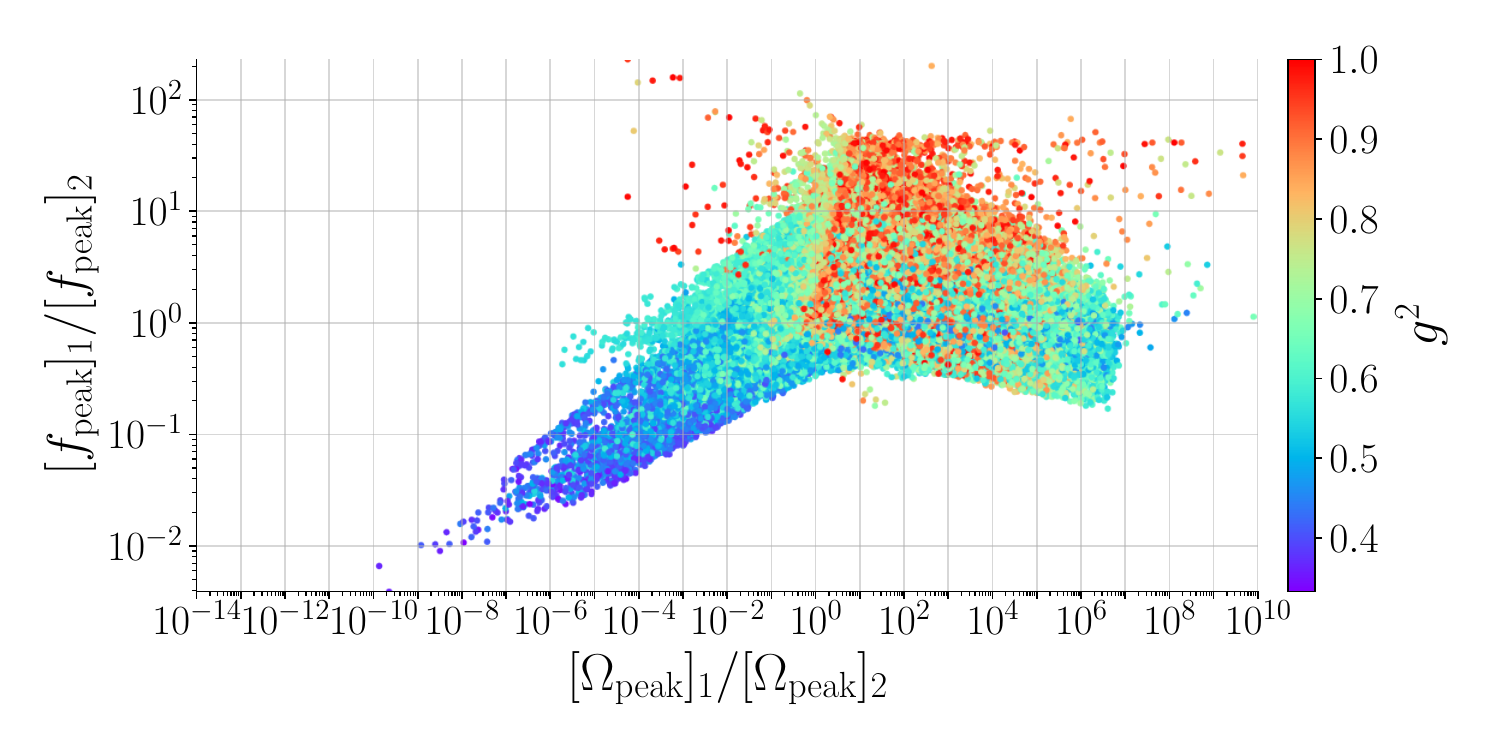} \\
    \caption{Scatter plots of the ratio of the peak frequencies as a function of the ratio of the corresponding peak amplitudes, for TM1 in the limit $g_1 = g_2 = g$. The color bar indicates 
    $\mathrm{log_{10}}(|\lambda_{12}|)$ in the top panel and $g$ in the bottom panel.   }
    \label{fig:scatter_equalg}
\end{figure}

We now turn our attention to the couplings $g_3$ and $y$, which are present in TM2 and TM3, respectively, but not in TM1. We do not find a strong correlation between these parameters and the thermodynamic and GW observables, although they play an important role in determining whether two-step transitions can occur. Similar to $\lambda_{12}$, both $g_3$ and $y$ induce mixing between $\varphi_1$ and $\varphi_2$.
which can prevent the formation of extrema along the individual field directions if these couplings are large. 
From Fig.~\ref{fig:scatter_y_g3}, it is evident that for $y,g_3 \gtrsim 0.1$ the allowed reheating temperature ratios are restricted to a narrow region close to the one-step transition limit, $[T_{\rm RH}]_1 \sim [T_{\rm RH}]_2$. Thus, we infer that $y,g_3 \gtrsim 0.1$ impedes the formation of intermediate minima at finite temperature and leads to one-step FOPTs. 
Conversely, for small values of these couplings, a broad range of temperature ratios is allowed, giving rise to spectra in which the two peak frequencies are well separated.

In Fig.~\ref{fig:scatter_equalg}, we show the same projections for TM1 as in the left panels of Fig.~\ref{fig:scan_TM1}, in the realistic limit of equal gauge couplings $g_1 = g_2 = g$. We find similar results for TM3. Note that the correlations between the spectral parameters and $g,\lambda_{12}$ are less pronounced compared to Fig.~\ref{fig:scan_TM1}. As before, larger values of $|\lambda_{12}|$ tend to prefer larger GW peak amplitude ratios (top panel), and larger $g^2$ correlates with larger peak frequency ratios (bottom panel). We therefore conclude that significantly different gauge couplings are not required for two-step FOPTs to occur. The sloping regions for $[\Omega_{\mathrm{peak}}]_1/[\Omega_{\mathrm{peak}}]_2 < 1$ in both panels arise because the suppression of the first peak (Eq.~\ref{eq:shift_rel}) is greater for points with large $\alpha_2$. The same feature is not present in Fig.~\ref{fig:scan_TM1} because the two gauge couplings produce a larger dispersion of color for small amplitude/frequency ratios. We now argue that for TM2, the equal gauge coupling limit $g = g_1 = g_2 = g_3$, is detrimental to the generation of two-step FOPTs. As discussed previously, small mixing parameters are required for two-step transitions to occur. On the other hand, successful nucleation and percolation require gauge couplings of $\mathcal{O}(0.1-1)$, which can induce sizable mixing through $g_3$. Indeed, from Fig.~\ref{fig:scatter_y_g3} we find that viable two-step transitions require $g_3 \lesssim 0.35$, while $g_1,g_2 \gtrsim 0.6$ is preferred. While it is possible to engineer a combination of $\lambda_{12}$ and $g_3$ that allows two-step transitions, in such cases one or both peaks have unobservably small amplitudes. We illustrate this  in the next section with a realistic model.

%%%%%%%%%%%%%%%%%%%%%%%%%%%%%%%%%%%%%%%%%%%%%%%%%%%%%%%%%%
\section{Realistic models}\label{sec:real_models}
%%%%%%%%%%%%%%%%%%%%%%%%%%%%%%%%%%%%%%%%%%%%%%%%%%%%%%%%%%
\begin{table*}[t]
	\centering
	\renewcommand{\arraystretch}{1.5}
	\resizebox{\textwidth}{!}{%
    \begin{tabular}{lcccccccccc}
		\hline
		& $g_X$ & $\lambda_h$ & $\lambda_{\Phi_1}$ & $\lambda_{\Phi_2}$ & $\lambda_3$ & $\lambda_4$ & $\lambda_{h1}$ & $\lambda_{h2}$ & $v_1~(\mathrm{GeV})$ & $v_2~(\mathrm{GeV})$ \\
		\hline
		& $0.25$ & $0.15$ & $-7.49\times 10^{-5}$ & $-5.78\times 10^{-5}$ & $6.61\times 10^{-5}$ & $1.47\times 10^{-4}$ & $1.37\times 10^{-3}$ & $-3.49\times 10^{-6}$ & $3455.08$ & $1519.66$ \\
		\hline
		& $\alpha_1$ & $\alpha_2$ & $[\beta/H(T_p)]_1$ & $[\beta/H(T_p)]_2$ & $[T_c]_1~(\mathrm{GeV})$ & $[T_c]_2~(\mathrm{GeV})$ & $[T_p]_1~(\mathrm{GeV})$ & $[T_p]_2~(\mathrm{GeV})$ & $[T_{\rm RH}]_1~(\mathrm{GeV})$ & $[T_{\rm RH}]_2~(\mathrm{GeV})$ \\
		\hline
		  & $1.63$ & $4.72$ & $17603.03$ & $19648.92$ & $170.69$ & $57.86$ & $58.12$ & $16.34$ & $73.99$ & $25.27$\\
		\hline
	\end{tabular}%
    }
	\caption{Model and thermodynamic parameters for the benchmark point of the dark SU(3) model. The corresponding GW spectrum is shown in Fig.~\ref{fig:spectra_SU3} with an SNR of 175.9 after 4 years of observation by BBO.}
	\label{tab:benchmarks_SU3}
\end{table*}

As stated earlier, realistic models can be mapped onto toy models. For example, a conformal model with a dark SU(3) gauge symmetry can be mapped onto TM2. Besides the SM Higgs doublet $\mathcal{H}$, the minimal realization contains two SU(3) triplets $\Phi_1$ and $\Phi_2$, both of which are singlets under the SM gauge group. We require the SM-like Higgs boson mass $M_h \in [ 125.0,125.5]~\mathrm{GeV}$. For a complete discussion of the model that incorporates the neutrino sector, and its GW phenomenology, see Ref.~\cite{Araujo:2026abc}. The tree-level potential of this model is~\cite{Karam:2016rsz,Araujo:2026abc}
\begin{equation}
\begin{split}
    V_0 =~& \lambda_{\Phi_1}(\Phi_1^\dagger \Phi_1)^2 
    + \lambda_{\Phi_2}(\Phi_2^\dagger \Phi_2)^2 
    - \lambda_3(\Phi_1^\dagger \Phi_1)(\Phi_2^\dagger \Phi_2) \\ 
    &+\lambda_4(\Phi_1^\dagger \Phi_2)(\Phi_2^\dagger \Phi_1) + \Big( \frac{\lambda_5}{2} (\Phi_1^\dagger \Phi_2)^2 \\
    &+\lambda_6(\Phi_1^\dagger \Phi_1)(\Phi_2^\dagger \Phi_1) 
    + \lambda_7 (\Phi_2^\dagger \Phi_2)(\Phi_1^\dagger \Phi_2) 
    + \text{h.c.} \Big) \\
    %\Big(\frac{\lambda_5}{2} (\Phi_1^\dagger \Phi_2)^2 + \text{h.c.} \Big) \nonumber\\
    &- \lambda_{h1}(\mathcal{H}^\dagger\mathcal{H})(\Phi_1^\dagger\Phi_1) + \lambda_{h2}(\mathcal{H}^\dagger\mathcal{H})(\Phi_2^\dagger\Phi_2)  \\
    &+ \left( \lambda_{h12}(\mathcal{H}^\dagger \mathcal{H})(\Phi_1^\dagger \Phi_2) + \text{h.c.} \right) + \lambda_h (\mathcal{H}^\dagger \mathcal{H})^2 \,.
    \end{split}
\end{equation}
To reduce the number of free parameters, we impose a $\mathbb{Z}_3$ symmetry $\Phi_i \rightarrow S_i \Phi_i$ (with $S_i = (1,e^{i2\pi/3})$) so that $\lambda_5,\, \lambda_6, \,\lambda_7,\,\lambda_{h12} = 0$. The scalar field masses are given by
\begin{equation}\label{eq:mass_mh}
\begin{split}
    &M_{h}^2 = \frac{1}{2}(A - \sqrt{A^2+B})\,, \\
    &M_{\phi_1}^2 = 0\,,\\
    &M_{\phi_2}^2 = \frac{1}{2}(A + \sqrt{A^2+B})\,,\\
    &M_{\phi_3}^2 = M_{\phi_4}^2 = \frac{1}{2}v_1^2 \lambda_4\,,% - \lambda_5
    %M_{\phi_4}^2 &= \frac{1}{2}v_1^2 \lambda_4\,, \label{eq:mass_mpsi4} %\lambda_5
        \end{split}
\end{equation}
where 
\begin{equation}
\begin{split}
   A = \frac{1}{2}[v_1^2 (\lambda_3 + \lambda_{h1}) + v_2^2 (\lambda_3 - \lambda_{h2})+ v_h^2 (\lambda_{h1} - \lambda_{h2})]\,, \\
    B = 4(v_h^2 + v_1^2 + v_2^2) [-v_1^2 \lambda_3 \lambda_{h1} + (v_2^2 \lambda_3 + v_h^2 \lambda_{h1}) \lambda_{h2}]\,.\nonumber
    \end{split}
\end{equation}
Here, $v_i$ is the vev of $\Phi_i$ and $v_h \simeq 246$~GeV is the Higgs vev. $\phi_1$ is the pseudo-Goldstone boson of the scale symmetry, and acquires mass from radiative corrections. The dark gauge bosons masses are
\begin{equation}\label{eq:gauge_masses}
\begin{split}
    &M_{X^{1,2}_\mu}^2 = \frac{1}{4} g_X^2 v_2^2\,, \\
    &M_{X^{3'}_\mu}^2 = \frac{1}{6} g_X^2 \left(v_1^2 + v_2^2 - \sqrt{v_1^4 - v_1^2 v_2^2 + v_2^4}\right)\,, \\
    &M_{X^{8'}_\mu}^2 = \frac{1}{6} g_X^2 \left(v_1^2 + v_2^2 + \sqrt{v_1^4 - v_1^2 v_2^2 + v_2^4}\right)\,, \\
    &M_{X^{4,5}_\mu}^2 = \frac{1}{4} g_X^2 v_1^2\,, \\
    &M_{X^{6,7}_\mu}^2 = \frac{1}{4} g_X^2 (v_1^2 + v_2^2)\,.
\end{split}
\end{equation}
\begin{figure}[t]
    \centering
    \includegraphics[width=0.5\textwidth]{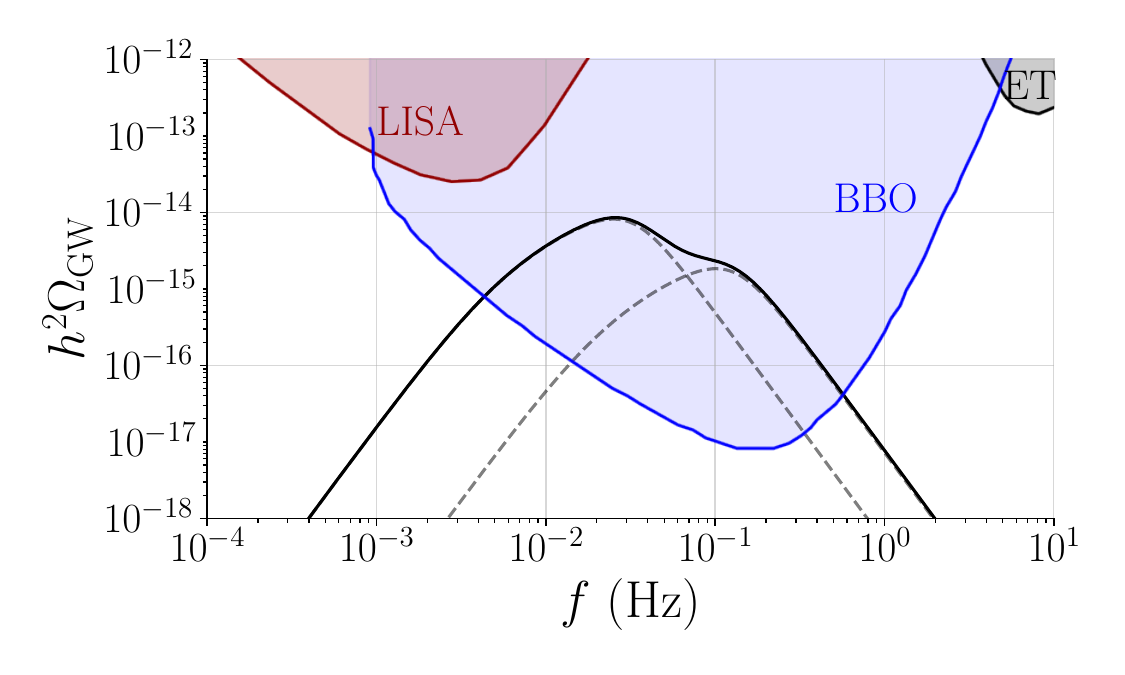} 
    \caption{Two-peak GW spectrum for the benchmark point of the dark SU(3) model in Table~\ref{tab:benchmarks_SU3}. We also show the BBO~\cite{Yagi:2011wg} sensitivity curve. }
    \label{fig:spectra_SU3}
\end{figure}

Due to the nonlinear nature of some of the scalar and dark gauge boson masses, we cannot make use of the previous approximations to write the effective potential in polynomial form, and instead employ \texttt{CosmoTransitions}. Furthermore,  we consider the full renormalization-group (RG) evolution of the couplings in the scalar potential, whose equations are provided in Appendix~\ref{app:rges}. The $\lambda_3$, $\lambda_4$, and $\lambda_5$ couplings in the SU(3) model play the role of $\lambda_{12}$ in TM2. Of these three couplings, we find that $\lambda_3$ is the main determinant of whether the transition exhibits a one-peak or two-peak structure. 
The gauge couplings $g_1$, $g_2$ and $g_3$ in TM2 are proportional to $g_X$ if the square-root terms in Eq.~\eqref{eq:gauge_masses} are neglected.
Note that $g_X$ affects the the peak structure by inducing mixing between $\Phi_1$ and $\Phi_2$. 

Using insights from TM2, with small mixing $(\lambda_3,\, \lambda_4 \lesssim 10^{-4})$ and a moderate size gauge coupling $(g_X \lesssim 0.6)$, we find two-peak GW spectra in the $\mathrm{SU(3)}$ model. While we do not find points for which both peaks can be detected by LISA or ET, we identify a benchmark point with both peaks detectable at the Big Bang Observer (BBO) in Table~\ref{tab:benchmarks_SU3}.
This point produces short and mildly supercooled phase transitions with large values of $\beta/H(T_p)$ and $\alpha = 1.63$ for the first transition, and $\alpha = 4.72$ for the second. The two-peak GW signal in Fig.~\ref{fig:spectra_SU3} is easily observable at BBO, with $\mathrm{SNR}_{\rm BBO} = 175.91$.

As for realistic models that map to TM1/TM3, dark U(1)$'$~\cite{Jinno:2016knw,Jaeckel:2016jlh,Ellis:2020nnr} and SU(2) models~\cite{Hambye:2018qjv,Marfatia:2020bcs,Kierkla:2022odc} do not generate two-step transitions since the transition along the Higgs direction proceeds as a smooth crossover. 
$\mathrm{U(1)'\times U(1)'}$ model constructions allow for a direct mapping to TM1/TM3 if each $\mathrm{U(1)'}$ corresponds to a different gauge symmetry (and therefore a different gauge coupling). Single $\mathrm{U(1)'}$ models with the addition of an extra complex singlet charged under $\mathrm{U(1)'}$ may also be viable. 

%%%%%%%%%%%%%%%%%%%%%%%%%%%%%%%%%%%%%%%%%%%%%%%%%%%%%%%%%%
\section{Summary}\label{sec:summ}
%%%%%%%%%%%%%%%%%%%%%%%%%%%%%%%%%%%%%%%%%%%%%%%%%%%%%%%%%%
We studied two-field conformal models that produce supercooled two-step phase transitions and characteristic two-peak GW spectra. Our goal was to study genuine two-peak spectra that result from two-field dynamics, as opposed to two-peak spectra obtained from independent single-field potentials.
%that can be tested for at future space- and ground-based interferometers such as LISA and ET. 
Using a set of toy models to capture the main features of more realistic models, and a 
semianalytical treatment of the thermal potential and the Euclidean action, we derived the necessary conditions to produce a two-step transition. Our results indicate that large values of the mixing parameters ($\lambda_{12}$, $g_3$ and $y$) strongly impede the generation of two-step transitions by reducing the duration for which an intermediate phase can exist. For small mixings ($y \lesssim 0.1$, $g_3 \lesssim 0.35$ and $|\lambda_{12}| \lesssim 0.01$), two-step transitions are possible and the  separation of the two peak frequencies is primarily controlled by the product of the gauge couplings $g_1g_2$. Finally, we demonstrated that these features arise in a realistic conformal dark SU(3) model, although achieving two simultaneously strong signals is challenging due to the mixing terms in the gauge masses. On the other hand, Abelian 
$\mathrm{U(1)'}$ or $\mathrm{U(1)'\times U(1)'}$ models have the necessary ingredients to produce observable two-peak spectra.

\begin{acknowledgments}
We thank M.~Zandi for discussions. J.G. and D.M. thank CERN for its hospitality while this work was in progress.  D.A. and J.G. are supported by LIP - Laboratório de Instrumentação e Física Experimental de Partículas under the Portuguese Foundation for Science and Technology (FCT - Funda\c{c}\~{a}o para a Ci\^{e}ncia e a Tecnologia) Multi-Annual Financing Program for R\&D Units with reference UID/50007/2025 (\url{https://doi.org/10.54499/UID/50007/2025}), and by LIP and FCT, reference LA/P/0016/2020 (\url{https://doi.org/10.54499/LA/P/0016/2020}). D.A. is also supported by a research grant EURAXESS 454862, within the scope of the Multiannual Funding of R\&D Units with reference UID/5007/2025.
D.M. is supported in part by the U.S. Department of Energy under Grant No. DE-SC0010504.
\end{acknowledgments}

\appendix
\section{Tadpole equations}
\label{app:tadpoles}

The tadpole equations for TM1 are
\begin{align}
    &\lambda_1 = \frac{1}{16}\Bigg[ \frac{g_1^4}{\pi^2} - \frac{8\lambda_{12}v_2^2}{v_1^2} - \frac{3g_1^4}{\pi^2}\ln\frac{g_1v_1^2}{\mu^2}\Bigg]\,, \label{eq:lambs_TM1_1}\\
    &\lambda_2 = \frac{1}{16}\Bigg[ \frac{g_2^4}{\pi^2} - \frac{8\lambda_{12}v_1^2}{v_2^2} - \frac{3g_2^4}{\pi^2}  \ln\frac{g_2v_2^2}{\mu^2}\Bigg]\,.\label{eq:lambs_TM2_2}
\end{align}
The tadpole equations for TM2 are
\begin{align}
    &\lambda_1 = \frac{1}{16 \pi ^2 v_1^2}\Bigg[v_1^2 \left(g_1^4+g_3^4\right)-3 g_1^4 v_1^2 \ln \left(\frac{g_1^2 v_1^2}{\mu^2}\right)  \nonumber\\
    &\hspace{2.5em} - 3 g_3^4 \left(v_1^2+v_2^2\right) \ln \left(\frac{g_3^2 \left(v_1^2+v_2^2\right)}{\mu ^2}\right)  \nonumber\\
    &\hspace{2.5em} + v_2^2 \left(g_3^4-8 \pi ^2 \lambda_{12}\right)\Bigg]\,,\\
    &\lambda_2 = \frac{1}{16 \pi ^2 v_2^2}\Bigg[v_2^2 \left(g_2^4+g_3^4\right)-3 g_2^4v_2^2 \ln \left(\frac{g_2^2 v_2^2}{\mu^2}\right)  \nonumber\\
    &\hspace{2.5em}- 3 g_3^4 \left(v_1^2+v_2^2\right) \ln \left(\frac{g_3^2 \left(v_1^2+v_2^2\right)}{\mu ^2}\right)  \nonumber\\
    &\hspace{2.5em} + v_1^2 \left(g_3^4-8 \pi ^2 \lambda_{12}\right)\Bigg]\,.
\end{align}
The tadpole equations for TM3 are
\begin{align}
    &\lambda_1 = \frac{1}{16 \pi ^2 v_1^3}\Bigg[g_1^4 v_1^3-3 g_1^4 v_1^3 \ln \left(\frac{g_1^2 v_1^2}{\mu ^2}\right)  \nonumber \\
    &\hspace*{2.5em} - 8 \pi ^2 \lambda_{12} v_1 v_2^2 - 2 y^4 (v_1+v_2)^3 \nonumber \\
    &\hspace*{2.5em} + 2 y^4 (v_1+v_2)^3 \ln \left(\frac{y^2 (v_1+v_2)^2}{\mu ^2}\right)\Bigg]\,, \label{eq:lambs_TM3_1} \\
    &\lambda_2 = \frac{1}{16 \pi ^2 v_2^3} \Bigg[g_2^4 v_2^3-3 g_2^4 v_2^3 \ln \left(\frac{g_2^2 v_2^2}{\mu ^2}\right)  \nonumber \\
    &\hspace*{2.5em} - 8 \pi ^2 \lambda_{12} v_1^2 v_2-2 y^4 (v_1+v_2)^3  \nonumber \\
    &\hspace*{2.5em} + 2 y^4 (v_1+v_2)^3 \ln \left(\frac{y^2 (v_1+v_2)^2}{\mu ^2}\right) \Bigg] \label{eq:lambs_TM3_2}\,.
\end{align}

\section{Polynomial expansion of the thermal potentials}\label{sec:poly_exp}

For TM2, the potential for the $(0,0)\rightarrow (v_1,0)$ transition is given by 
\begin{align}
    & V_0(T) = -\left(5 C_b^{3/2} \left(g_1^3+g_3^3\right)+6 \pi ^3\right)\frac{T^4}{60\pi}  \,,\label{eq:V0_T_TM2} \\
    & m^2(T) =\left(3 \pi  \left(g_1^2+g_3^2\right) - 2 C_b^{3/2} c_2 \left(g_1^3+g_3^3\right)\right) \frac{T^2}{12\pi}\,, \label{eq:m2T_TM2} \\
    & K(T) = \left(C_b^{3/2} c_3+2\right) \left(g_1^3+g_3^3\right) \frac{T}{4\pi}\,, \label{eq:KT_TM2} \\
    & \lambda(T) = \frac{1}{8\pi^2} \Big[g_1^4 (3 \gamma_E-1-6 \ln2) - 3 \left(g_1^4+g_3^4\right) \ln(\pi T/\mu)\Big]  \nonumber\\
    &\hspace*{3em} + \frac{1}{8\pi^2}\Big[g_3^4 (3 \gamma_E-1-6 \ln2)-8 \pi ^2 \lambda_1\Big] 
    \nonumber\\ &\hspace*{3em} 
   + \frac{C_b^{3/2}c_4}{3 \pi } \left(g_1^3+g_3^3\right)
   \,.\label{eq:lambT_TM2}
\end{align}
The potential for the $(0,0)\rightarrow (0,v_2)$ transition is obtained by the replacements, $g_1 \rightarrow g_2$ and $\lambda_1 \rightarrow \lambda_2$ in Eqs.~\eqref{eq:V0_T_TM2}--\eqref{eq:lambT_TM2}. For a  transition in the $(v_1,0)\rightarrow (v_1,v_2)$ direction, 
\begin{align}
    & V_0(T) = \frac{3v_1^4}{32\pi^2}  \left(g_1^4+g_3^4\right) \ln (4 \pi  T/\mu)\nonumber\\
    &\hspace*{3em} + \frac{v_1^4}{32 \pi ^2} \left(8 \pi ^2 \lambda_1-3 \gamma_E \left(g_1^4+g_3^4\right)\right) \nonumber\\
    &\hspace*{3em} + \frac{1}{40} \left(5 v_1^2 \left(g_1^2+g_3^2\right) T^2-4 \pi ^2 T^4\right)  \nonumber\\
    &\hspace*{3em} + \frac{1}{32\pi^2}\Big[v_1^4 \left(g_1^4+g_3^4\right)-8 \pi g_1^3 v_1^3 T-8 \pi  g_3^3 v_1^3 T\Big] \nonumber\\
    &\hspace*{3em}-  C_b^{3/2}  \left(g_2^3+g_3^3\right) \frac{T^4}{12\pi} + V_1(v_1,0)\,,
    \label{eq:V0_T_TM2_newdir} \\
    & m^2(T) = \frac{1}{4}\left[\left(g_2^2+g_3^2\right) T^2+2 \lambda_{12} v_1^2\right] \nonumber\\
    &\hspace*{3em} + \frac{1}{8\pi^2} \Big[g_3^4 v_1^2 (-3
   \gamma_E +3 \ln(\pi T/\mu)+1+\ln 64)\Big]  \nonumber\\
    &\hspace*{3em} - \frac{6 g_3^3 v_1 T}{8 \pi} -C_b^{3/2}c_2  \left(g_2^3+g_3^3\right)\frac{T^2}{6 \pi}\,,
   \label{eq:m2T_TM2_newdir} \\
    & K(T) =\Big[\left(g_2^3 \left(C_b^{3/2}c_3+2\right)+g_3^3 \left( C_b^{3/2} c_3-1\right)\right)\Big] \frac{T}{4\pi}\,, \label{eq:KT_TM2_newdir} \\
    & \lambda(T) = \frac{1}{16} \Bigg( \frac{6 g_3^3 T}{\pi v_1} -16 \lambda_2 - \frac{6}{\pi^2} \left(g_2^4+g_3^4\right) \ln(\pi T/\mu) \nonumber\\
    &\hspace*{3em} + \frac{1}{\pi^2}\Big[(6 \gamma_E-2-3 \ln16) \left(g_2^4+g_3^4\right)\Big]  \nonumber\\
    &\hspace*{3em}  + \frac{C_b^{3/2}c_4 }{3 \pi }\left(g_2^3+g_3^3\right)\Bigg)\,,\label{eq:lambT_TM2_newdir}
\end{align}
where
\begin{align}\label{eq:vdaisy_cont}
    & V_1(v_1,0) = -\frac{T}{48\pi}\Big[\Big(-4 g_1^3 v_1^3+\sqrt{2} \left(g_1^2 \left(3 T^2+2 v_1^2\right)\right)^{3/2} \nonumber \\
    &\hspace*{5em} + 3 \sqrt{6} g_2^3 T^3-4 g_3^3 v_1^3 \nonumber\\
    &\hspace*{5em} +\sqrt{2} \left(g_3^2 \left(3 T^2+2 v_1^2\right)\right)^{3/2}\Big)\Big]\,.
\end{align}
The potential for the $(0,v_2)\rightarrow (v_1,v_2)$ transition is obtained by the replacements, $g_1 \leftrightarrow g_2$,  $\lambda_1 \leftrightarrow \lambda_2$ and $v_1 \rightarrow v_2$ in Eqs.~\eqref{eq:V0_T_TM2_newdir}--\eqref{eq:vdaisy_cont}. 

For  TM3, along the $(0,0)\rightarrow (v_1,0)$ direction we write
\begin{align}
    & V_0(T) = - \left(30 C_b^{3/2} g_1^3+31 \pi ^3\right)\frac{T^4}{360 \pi } \,, \label{eq:V0_T_TM3} \\
    & m^2(T) = \left(\pi  \left(3 g_1^2+y^2\right)-2 C_b^{3/2} c_2g_1^3\right)  \frac{T^2}{12 \pi }\,, \label{eq:m2T_TM3} \\
    & K(T) =\left(C_b^{3/2}c_3+2\right)  \frac{g_1^3 T}{4 \pi }\,, \label{eq:KT_TM3} \\
    & \lambda(T) = -\frac{1}{8\pi^2}\Big[-3 \gamma_E g_1^4+g_1^4+8 \pi ^2 \lambda_1+2 \gamma_E y^4\Big]  \nonumber \\
    &\hspace*{3em} - \frac{1}{8\pi^2}\Big[\left(3 g_1^4-2 y^4\right) \ln(\pi T/\mu) +g_1^4 \ln 64\Big] \nonumber \\
    &\hspace*{3em} + \frac{C_b^{3/2} c_4 g_1^3}{3 \pi }\,.\label{eq:lambT_TM3}
\end{align}
The potential for the $(0,0)\rightarrow (0,v_2)$ transition is obtained by the replacements, $g_1 \rightarrow g_2$ and $\lambda_1 \rightarrow \lambda_2$ in Eqs.~\eqref{eq:V0_T_TM3}--\eqref{eq:lambT_TM3}. For a transition in the  $(v_1,0)\rightarrow (v_1,v_2)$ direction,
\begin{align}
    & V_0(T) = \frac{v_1^4}{32\pi^2}\Big[\left(\left(3 g_1^4-2 y^4\right) \ln(\pi T/\mu)+g_1^4 \ln64\right)\Big]  \nonumber \\
    &\hspace*{3em} + \frac{v_1^4}{32\pi^2}\Big[\left(-3 \gamma_E g_1^4+8 \pi ^2 \lambda_1+2 \gamma_E y^4\right)\Big]  \nonumber \\
    &\hspace*{3em} + \frac{g_1^2 v_1^2}{32\pi^2}\Big[\left(g_1^2 v_1^2-8 \pi  g_1 v_1 T+4 \pi ^2 T^2\right)\Big] \nonumber \\
    &\hspace*{3em} + \frac{1}{360} \left(15v_1^2 y^2 T^2-31 \pi ^2 T^4\right) - C_b^{3/2} g_2^3 \frac{T^4}{12 \pi } \nonumber\\
    &\hspace*{3em} + V_{\rm Daisy}(v_1,0)\,,
    \label{eq:V0_T_TM3_newdir} \\
    & m^2(T) = \frac{1}{12} \Bigg(\left(3 g_2^2+y^2\right)T^2-\frac{9 v_1^2 y^4}{\pi ^2}\ln(\pi T/\mu)  \nonumber \\
    &\hspace*{3em} + 6 \lambda_{12} v_1^2+\frac{9 \gamma_E v_1^2 y^4}{\pi ^2}\Bigg) - C_b^{3/2} c_2\frac{ g_2^3 T^2}{6 \pi }\,, \label{eq:m2T_TM3_newdir} \\
    & K(T) = \frac{3}{4\pi^2}\Big[\left(\pi  g_2^3 T+v_1 y^4 \ln(\pi T/\mu) - \gamma_E v_1 y^4\right)\Big]  \nonumber \\
    &\hspace*{3em} + \left(C_b^{3/2}c_3-1\right)\frac{g_2^3 T}{4 \pi} \,, \label{eq:KT_TM3_newdir} \\
    & \lambda(T) = \frac{1}{8\pi^2}\Big[g_2^4 \big[3 \gamma_E-1 - \ln(64 \pi ^3)\big]+3 g_2^4 \ln\mu - 8 \pi ^2
    \lambda_2\Big]  \nonumber \\
    &\hspace*{3em} - \frac{1}{8\pi^2} \Big[3 g_2^4 \ln T + 2 y^4 (\gamma_E - \ln(\pi T/\mu))\Big] 
    \nonumber \\
    &\hspace*{3em} + \frac{C_b^{3/2} c_4 g_2^3}{3 \pi }\,.\label{eq:lambT_TM3_newdir}
\end{align}
The potential for the $(0,v_2)\rightarrow (v_1,v_2)$ transition is obtained by the replacements, $g_1 \leftrightarrow g_2$,  $\lambda_1 \leftrightarrow \lambda_2$ and  $v_1 \rightarrow v_2$ in Eqs.~\eqref{eq:V0_T_TM3_newdir}--\eqref{eq:lambT_TM3_newdir}. 

\section{RGEs for the dark SU(3) model}\label{app:rges}

The beta functions for the SU(3) model,
\begin{equation*}
\beta\left(X\right) \equiv \mu \frac{d X}{d \mu}\equiv\frac{1}{\left(4 \pi\right)^{2}}\beta^{(1)}(X)\,.
\end{equation*}
are given by 
\begin{align}
& \beta^{(1)}(g_Y) =\frac{41}{6} g_Y^{3}\,, \\
& \beta^{(1)}(g_W) =- \frac{19}{6} g_W^{3}\,, \\
& \beta^{(1)}(g_S) =-7 g_S^{3}\,, \\
& \beta^{(1)}(g_X) =- \frac{32}{3} g_X^{3} -\frac{1}{(4\pi)^2}\left(\frac{284}{3}g_X^5\right)\,, \\
& \beta^{(1)}(y_t) = \frac{9}{2} y_t^3 -  \frac{17}{12} g_Y^{2} y_t -  \frac{9}{4} g_W^{2} y_t- 8 g_S^{2} y_t\,, \\
& \beta^{(1)}(\lambda_h) = 24 \lambda_h^{2} + 3 \lambda_{h1}^{2} + 3 \lambda_{h2}^{2} - 3 g_Y^{2} \lambda_h \nonumber\\
&\hspace*{5em} - 9 g_W^{2} \lambda_h + \frac{3}{8} g_Y^{4} + \frac{3}{4} g_W^{2} g_Y^{2} + \frac{9}{8} g_W^{4} \nonumber\\
&\hspace*{5em} + 12 \lambda_h y_t^2 - 6 y_t^4\,, \\
&\beta^{(1)}(\lambda_{\Phi_1}) = 28 \lambda_{\Phi_1}^{2} + 3 \lambda_3^{2} + 2 \lambda_3 \lambda_4 + \lambda_4^{2} + 2 \lambda_{h1}^{2} \nonumber\\
&\hspace*{5em} - 16 g_X^{2} \lambda_{\Phi_1} + \frac{13}{6} g_X^{4}\,,\\ % + \left|{\lambda_5}\right|^{2}
&\beta^{(1)}(\lambda_{\Phi_2}) = 28 \lambda_{\Phi_2}^{2} + 3 \lambda_3^{2} + 2 \lambda_3 \lambda_4 + \lambda_4^{2} + 2 \lambda_{h2}^{2} \nonumber\\
&\hspace*{5em} - 16 g_X^{2} \lambda_{\Phi_2} + \frac{13}{6} g_X^{4}\,,\\ %+ \left|{\lambda_5}\right|^{2}
&\beta^{(1)}(\lambda_3) = 16 \lambda_{\Phi_1} \lambda_3 + 4 \lambda_{\Phi_1} \lambda_4 + 16 \lambda_{\Phi_2} \lambda_3 + 4 \lambda_{\Phi_2} \lambda_4 \nonumber\\
&\hspace*{5em} + 4 \lambda_3^{2} + 2 \lambda_4^{2} + 4 \lambda_{h1} \lambda_{h2} \nonumber\\ %+ 2 \left|{\lambda_5}\right|^{2}
&\hspace*{5em} - 16 g_X^{2} \lambda_3 + \frac{11}{6} g_X^{4}\,, \\
&\beta^{(1)}(\lambda_4) = 4 \lambda_{\Phi_1} \lambda_4 + 4 \lambda_{\Phi_2} \lambda_4 + 8 \lambda_3 \lambda_4 + 6 \lambda_4^{2} \nonumber\\
&\hspace*{5em} - 16 g_X^{2} \lambda_4 + \frac{5}{2} g_X^{4}\,, \\ % + 10 \left|{\lambda_5}\right|^{2}
& \beta^{(1)}(\lambda_{h1}) = 12 \lambda_h \lambda_{h1} + 16 \lambda_{\Phi_1} \lambda_{h1} + 6 \lambda_3 \lambda_{h2} \nonumber\\
&\hspace*{5em} + 2 \lambda_4 \lambda_{h2} + 4 \lambda_{h1}^{2} - \frac{3}{2} g_Y^{2} \lambda_{h1} - \frac{9}{2} g_W^{2} \lambda_{h1} \nonumber\\
&\hspace*{5em} - 8 g_X^{2} \lambda_{h1} + 6 \lambda_{h1} y_t^2\,,\\
&\beta^{(1)}(\lambda_{h2}) = 12 \lambda_h \lambda_{h2} + 16 \lambda_{\Phi_2} \lambda_{h2} + 6 \lambda_3 \lambda_{h1} \nonumber\\
&\hspace*{5em} + 2 \lambda_4 \lambda_{h1} + 4 \lambda_{h2}^{2} - \frac{3}{2} g_Y^{2} \lambda_{h2} - \frac{9}{2} g_W^{2} \lambda_{h2} \nonumber\\
&\hspace*{5em} - 8 g_X^{2} \lambda_{h2} + 6 \lambda_{h2} y_t^2\,. %\\
%& \beta^{(1)}(\lambda_5) = 4 \lambda_{\Phi_1} \lambda_5 + 4 \lambda_{\Phi_2} \lambda_5 + 8 \lambda_3 \lambda_5 + 16 \lambda_4 \lambda_5 \nonumber\\
%&\hspace*{5em} - 16 g_X^{2} \lambda_5\,.
\end{align}
Here, $g_Y$, $g_W$, $g_S$ are the Standard Model hypercharge, weak, strong gauge couplings, respectively, and $y_t$ is the top Yukawa coupling.

\newpage
\bibliography{Refs.bib}

\end{document}